\documentclass[%
 reprint,
 amsmath,amssymb,
 aps,prl
]{revtex4-1}

\usepackage{graphicx}
\usepackage{dcolumn}
\usepackage{bm}
\usepackage{amsmath}

\begin{document}

\preprint{APS/123-QED}

\title{Contact network changes in ordered and disordered disk packings}

\author{Philip J. Tuckman}
\affiliation{Department of Physics, Yale University, New Haven, Connecticut, 06520, USA.}
\author{Kyle VanderWerf}
\affiliation{Department of Physics, Yale University, New Haven, Connecticut, 06520, USA.}
\author{Ye Yuan}
\email{yuanyepeking@pku.edu.cn}
\affiliation{Department of Mechanics and Engineering Science, College of Engineering, Peking University, Beijing 100871, China.}
\affiliation{Department of Mechanical Engineering and Materials Science, Yale University, New Haven, Connecticut 06520, USA.}

\author{Shiyun Zhang}
\email{zsy12@mail.ustc.edu.cn}
\affiliation{Department of Mechanics and Engineering Science, College of Engineering, Peking University, Beijing 100871, China.}
\affiliation{Department of Mechanical Engineering and Materials Science, Yale University, New Haven, Connecticut 06520, USA.}

\author{Jerry Zhang}
\affiliation{Department of Mechanical Engineering and Materials Science, Yale University, New Haven, Connecticut 06520, USA.}
\author{Mark D. Shattuck}
\affiliation{Benjamin Levich Institute and Physics Department, The City College of New York, New York, New York 10031, USA.}
\author{Corey S. O'Hern}
\email{corey.ohern@yale.edu}
\affiliation{Department of Physics, Yale University, New Haven, Connecticut, 06520, USA.}
\affiliation{Department of Applied Physics, Yale University, New Haven, Connecticut 06520, USA. Email: corey.ohern@yale.edu}
\affiliation{Department of Mechanical Engineering and Materials Science, Yale University, New Haven, Connecticut 06520, USA.}
\affiliation{Graduate Program in Computational Biology and Bioinformatics, Yale University, New Haven, Connecticut 06520, USA.}

\begin{abstract}
We investigate the mechanical response of packings of purely repulsive, frictionless disks to quasistatic deformations. The deformations include simple shear strain at constant packing fraction and at constant pressure, ``polydispersity'' strain (in which we change the particle size distribution) at constant packing fraction and at constant pressure, and isotropic compression.  For each deformation, we show that there are two classes of changes in the interparticle contact networks: jump changes and point changes. Jump changes occur when a contact network becomes mechanically unstable, particles ``rearrange'', and the potential energy (when the strain is applied at constant packing fraction) or enthalpy (when the strain is applied at constant pressure) and all derivatives are discontinuous. During point changes, a single contact is either added to or removed from the contact network. For repulsive linear spring interactions, second- and higher-order derivatives of the potential energy/enthalpy are discontinuous at a point change, while for Hertzian interactions, third- and higher-order derivatives of the potential energy/enthalpy are discontinuous. We illustrate the importance of point changes by studying the transition from a hexagonal crystal to a disordered crystal induced by applying polydispersity strain. During this transition, the system only undergoes point changes, with no jump changes. We emphasize that one must understand point changes, as well as jump changes, to predict the mechanical properties of jammed packings. 

\end{abstract}

\maketitle


\section{\label{sec:intro}Introduction}

Granular materials, which are composed of macroscopic grains that
interact via frictional contact forces, are ubiquitous in the natural
world and industrial applications. Unless they are continuously
driven, granular materials will come to rest and, when confined, they exist in jammed, solid-like states\cite{jamming_review}. The mechanical response of jammed granular materials is highly nonlinear, which gives rise to shear jamming\cite{shear_jamming}, intermittency and avalanches\cite{avalanches,avalanches2}, shear banding\cite{shear_banding1,shear_banding}, and other collective behavior\cite{collective}. 

Numerous theoretical and computational studies have focused on simplified descriptions of dry granular media, where they are modeled as packings of frictionless, purely repulsive spherical grains\cite{jamming,liu_review}. These studies have provided significant insights into the jamming transition in packings of frictionless, spherical particles.  Disordered packings of frictionless spherical particles are typically isostatic at jamming onset\cite{witten}, i.e. they possess the same number of interparticle contacts $N_c$ as the number of non-trivial degrees of freedom: $N_c = N_c^{\rm iso}$, where $N_c^{\rm iso}= dN - d + 1$ (for systems with periodic boundary conditions), $N$ is the number of (non-rattler\cite{rattler}) grains, and $d=2$, $3$ is the spatial dimension.  Ordered or compressed jammed packings can be hyperstatic with $N_c \ge N_c^{\rm iso}$\cite{hyperstatic}. Each jammed packing exists in a local energy minimum in configuration space, and therefore possesses a percolating network of non-zero interparticle forces and nonzero bulk and shear moduli. In contrast, packings with fewer contacts than the isostatic value, $N_c < N_c^{\rm iso}$, are unjammed and all interparticle forces are zero\cite{shen}.  Several studies have shown that isostatic jammed packings possess unique structural and mechanical properties, such as an excess number of low-frequency vibrational modes above the Debye prediction for the density of states\cite{wyart,silbert} and the power-law scaling of the shear modulus with increasing pressure\cite{goodrich}. 

In prior studies, we considered jammed packings of frictionless, spherical particles undergoing quasistatic deformation (i.e. steps of applied simple or pure shear strain with each step followed by energy minimization)\cite{sheng}.  During quasistatic deformation, grains in the packings undergo continuous motions along "geometric families," in which the network of interparticle contacts does not change\cite{families,families2}. The continuous geometric families are punctuated by particle rearrangements, which cause the contact networks to change. Such rearrangements determine the structural and mechanical properties of jammed packings. For example, particle rearrangements control the power-law scaling of the ensemble-averaged shear modulus as a function of pressure during isotropic compression\cite{kyle_prl}. Prior studies of sheared particulate materials have shown that there are two types of changes in the contact networks\cite{manning}. We refer to these contact network changes as 1) jump changes and 2) point changes. These previous studies also found that the relative frequency of jump and point changes is roughly constant with increasing system size.

In this work, we further investigate jump and point changes in the contact network and show that these two types of contact network changes occur during a wide range of quasistatic deformations in model granular materials. We carry out discrete element method simulations of purely repulsive, frictionless disks in 2D, focusing on several types of quasistatic deformations: simple shear strain, changes in the size polydispersity of the grains, and isotropic compression. For jump changes, jammed packings become mechanically unstable during quasistatic deformation\cite{lacks}, the particles rearrange, and as a result, the total energy, pressure, shear stress, and other thermodynamic quantities are discontinuous at the strain where the particle rearrangement occurs\cite{rearrangement}.  At a point change, a contact is added or removed from the interparticle contact network at a given strain, but the particles do not move significantly.  The positions of the particles are continuous with strain, but the derivatives of the particle positions with respect to strain are discontinuous. As a result, for point changes, the potential energy (in the case of strain applied at fixed packing fraction) or enthalpy (in the case of strain applied at fixed pressure) and their first derivatives are continuous as a function of strain\cite{cutoff}. For repulsive linear spring interactions, second- and higher-order derivatives of the potential energy/enthalpy are discontinuous at a point change, while for Hertzian interactions, third- and higher-order derivatives of the potential energy/enthalpy are discontinuous. We illustrate the importance of point changes by starting with a perfectly ordered jammed disk packing, adding small increments of size polydisperity to the system, and minimizing the potential energy (at fixed packing fraction) or enthalpy (at fixed pressure).  This system undergoes a series of point changes as it proceeds from a hyperstatic toward an isostatic state\cite{disordered_crystal1,disordered_crystal}. 

The remainder of the article is organized as follows. In Sec.~\ref{methods}, we describe the numerical methods that we use to generate disk packings at jamming onset and that we use to deform the jammed packings. In Sec.~\ref{results}, we show results for the coordination number ($z=2N_c/N$), total potential energy, shear stress, pressure, and other thermodynamic properties of jammed packings as a function of strain for each type of deformation, which allows us to illustrate point and jump changes. These studies are performed for both ordered packings of monodisperse disks and disordered packings of polydisperse disks. 
In Sec.~\ref{conclusions}, we summarize the conclusions and provide several possible future research directions including determining how point and jump changes separately contribute to the power-law scaling of the shear modulus with pressure during isotropic compression and investigating the effects of point changes in disk packings that interact via repulsive Hertzian spring interactions\cite{hertzian} and in jammed systems containing frictional and non-spherical particles. 

\section{Methods}
\label{methods}

We consider packings of $N$ circular disks in rectangular cells with area $A=L_x L_y$ and periodic boundary conditions in both the $x$- and $y$-directions. We study packings of monodisperse disks, for which there is significant positional order, as well as disordered packings of polydisperse disks.  The monodisperse disk packings possess jammed packing fractions near the value for the hexagonal lattice, $\phi_x = 0.907$, whereas the disordered polydisperse disk packings possess jammed packing fractions $\phi_J \approx 0.81$-$0.84$.   

The disks interact via the following purely repulsive pair potential: 
\begin{equation}
\label{spring}
    U(r_{ij}) = \frac{\epsilon}{\alpha} \left(1-\frac{r_{ij}}{\sigma_{ij}} \right)^{\alpha} \Theta \left( 1-\frac{r_{ij}}{\sigma_{ij}} \right),
\end{equation}
where $\epsilon$ is the characteristic energy scale of the repulsive interaction potential, the exponent of the interaction potential $\alpha=2$ for repulsive linear springs and $\alpha =5/2$ for "Hertzian" springs, $r_{ij}$ is the center-to-center distance between disks $i$ and $j$, $\hat{r}_{ij}=\vec{r}_{ij} / r_{ij}$, $\sigma_{ij}=(\sigma_i + \sigma_j)/2$ is the average diameter of disks $i$ and $j$, and the Heaviside function $\Theta(\cdot)$ ensures that the interaction is nonzero only when the disks overlap ($r_{ij} < \sigma_{ij}$).  The total potential energy is given by $U=\sum^N_{i=1} \sum^N_{j>i} U(r_{ij})$.  The repulsive force on disk $i$, arising from an overlap with disk $j$, is $\vec{F}(r_{ij})= {\vec \nabla}_{r_{ij}} U = \frac{\epsilon}{\sigma_{ij}} \left(1-\frac{r_{ij}}{\sigma_{ij}} \right)^{\alpha-1} \Theta \left( 1-\frac{r_{ij}}{\sigma_{ij}} \right) \hat{r}_{ij}$. Studies have shown that disks interacting via the purely repulsive potential in Eq.~\ref{spring} recapitulate the structural and mechanical properties of hard-sphere systems near jamming onset\cite{unified}.   

Note that the Hertzian theory for the force between two contacting elastic spherical particles depends on the spatial dimension.  The theory gives an exponent of $\alpha=5/2$ for the interaction energy between two elastic spheres in 3D and an exponent of $\alpha=2$ for the interaction between two parallel cylinders~\cite{johnson}, which can mimic interactions between elastic disks in 2D.  Thus, formally, ``Hertzian" interactions between elastic disks should consider $\alpha=2$ in 2D, not $\alpha=5/2$.  However, our goal was to investigate the effect of variations of the power-law exponent in Eq.~\ref{spring} on contact changes. Thus, we study both $\alpha=2$ and $5/2$ for disk packings in 2D, and refer to the $5/2$ exponent as the ``Hertzian" value since this is value of the exponent in 3D~\cite{jamming}.  

To generate jammed packings, we first randomly place $N$ disks in the simulation cell at small packing fraction $\phi_0 \approx 0.1$. We set the particle diameters to be $\sigma_i = \langle \sigma \rangle+\eta \delta_i$, where $-0.5 \le \delta_i/\langle \sigma \rangle \le 0.5$ is uniformly distributed, $\langle \delta_i \rangle = 0$, $\eta \langle \sigma \rangle /\sqrt{12}$ is the standard deviation of the disk diameters, and $\langle \sigma \rangle = N^{-1} \sum_{i=1}^N \sigma_i$ defines the average diameter. For disordered packings, we employ a square box, whereas for crystalline packings, we employ a rectangular box with aspect ratio $\sqrt{3}/2$, which allows a hexagonal packing of contacting disks to fit in the simulation cell without any defects. We isotropically compress the system in small packing fraction steps, $\Delta \phi$, until the system develops a small nonzero pressure, $p= A^{-1} \sum^N_{i=1} \sum^N_{j>i} {\vec f}_{ij} \cdot {\vec r}_{ij} > 0$. After each compression step, the total potential energy is minimized using the FIRE algorithm\cite{FIRE} until the magnitude of the total net force on the disks, $\sum_{i=1}^N |{\vec f}_i| < 10^{-14}$.  We study the coordination number, total potential energy, pressure, shear stress, and elastic moduli in jammed packings as a function of the packing fraction and strain. We measure energy, stress, and force in units of $\epsilon$, $\epsilon/\langle \sigma \rangle^2$, and $\epsilon/\langle \sigma \rangle$, respectively.  

To understand the effects of jump and point changes in the interparticle contact networks, we consider jammed disk packings undergoing several types of quasistatic deformations: 1) simple shear at constant packing fraction, 2) simple shear at constant pressure, 3) increments of increasing size polydispersity at constant packing fraction, 4) increments of increasing size polydispersity at constant pressure, and 5) isotropic compression. 

\subsection{Simple shear strain at fixed packing fraction}
\label{simple}

For simple shear deformations, the particle positions are transformed to $(x_i',y_i') = (x^0_i + \gamma L_x y^0_i/L_y,y^0_i)$ consistent with Lees-Edwards boundary conditions, where $(x^0_i,y^0_i)$ are the initial particle positions. After each simple shear strain step $\gamma$, we minimize the total potential energy at constant packing fraction until the system is in force balance, such that $\sum_{i=1}^N |{\vec f}_i| < 10^{-14}$.

During the simple shear strain deformation, we calculate several quantities as 
a function of $\gamma$ including the shear stress,
\begin{equation}
    \Sigma_{\gamma} = -\frac{1}{A} \frac{dU}{d\gamma} = -\frac{1}{L_y^2}  \sum^N_{i=1} \sum^N_{j>i} F_{yij} x_{ij},
\end{equation}
which becomes
\begin{equation}
    \Sigma_{\gamma} = \frac{\epsilon}{L_y^2}  \sum^N_{i=1} \sum^N_{j>i}  \frac{x_{ij} y_{ij}}{r_{ij} \sigma_{ij}} \left(1-\frac{r_{ij}}{\sigma_{ij}} \right) \Theta \left( 1-\frac{r_{ij}}{\sigma_{ij}} \right) 
\end{equation}
for repulsive linear spring interactions ($\alpha=2$ in Eq.\ref{spring}) (where $y_{ij}=y_i-y_j$, $x_{ij}=x_i-x_j$, and $dx_{ij}/d\gamma = y_{ij} L_x/L_y$)\cite{Lemaitre}, and 
the shear modulus,
\begin{equation}
\label{mod_def}
    G_{\gamma} \equiv -\frac{d\Sigma_{\gamma}}{d\gamma}.
\end{equation}
The shear modulus can be decomposed into the affine and nonaffine contributions\cite{mizuno}, $G_{\gamma}=G^a_{\gamma} + G^{na}_{\gamma}$, respectively. To calculate $G^a_{\gamma}$, we assume that all particles move according to the affine deformation, $(x_i',y_i') = (x^0_i + \gamma L_x y^0_i/L_y,y^0_i)$.  $G^{na}_{\gamma}$ includes the nonaffine particle motion in response to potential energy minimization at fixed packing fraction and boundary strain. For repulsive linear spring interactions ($\alpha=2$ in Eq.\ref{spring}), the affine contribution to the shear modulus can be calculated analytically, 
\begin{equation}
\label{shear_modulus}
G^a_{\gamma} = \epsilon \frac{L_x}{L_y^3} \sum^N_{i=1} \sum^N_{j>i} \left( \frac{x_{ij}^2 y_{ij}^2}{\sigma_{ij} r_{ij}^3} - \frac{y_{ij}^2}{\sigma_{ij} r_{ij}} \left(1-\frac{r_{ij}}{\sigma_{ij}} \right)\right) \Theta \left( 1-\frac{r_{ij}}{\sigma_{ij}} \right).
\end{equation}
We monitor $U$, $\Sigma_{\gamma}$, $G_{\gamma}$, and $G^a_{\gamma}$ before and after jump and point changes during the applied simple shear strain. 

\subsection{Simple shear strain at fixed pressure}
\label{shear_pressure}
We also apply quasistatic simple shear strain as described in Sec.~\ref{simple}, except at constant pressure. At each strain increment, we set the target pressure $p_t$ and minimize the enthalpy, $H=U+p_t A$. After each strain step, we terminate the minimization when $\sum_{i=1}^N  | {\vec \nabla}_{ {\vec r}_i,L_x} H| < 10^{-13}$. Minimizing the enthalpy ensures that we can maintain constant pressure $p_t$ as the system is strained. At each strain step, we measure the enthalpy and its derivative
$dH/d\gamma$ with respect to shear strain, and monitor jump and point changes in the interparticle contact network. 

\subsection{Polydispersity strain at fixed packing fraction}
\label{poly_phi}

In this section, we describe simulations in which we start the system with monodisperse ($\eta = 0$) or polydisperse disks ($\eta \langle \sigma \rangle =0.08$), and increase $\eta$ in small steps $\Delta \eta  \sim 10^{-5}$ to increase the polydispersity of the disks. After each increment, $\Delta \eta$, we reset the packing fraction to its desired value and minimize the total potential energy at constant packing fraction.  We measure the "polydispersity stress" as a function of $\eta$,
\begin{equation}
\Sigma_{\eta} = -\frac{1}{A} \frac{dU}{d\eta},
\end{equation}
which becomes 
\begin{equation}
\Sigma_{\eta} = -\frac{\epsilon}{L_x L_y} \sum^N_{i=1} \sum^N_{j>i} \left(1-\frac{r}{\sigma_{ij}} \right) \frac{r_{ij}}{\sigma_{ij}^2} \frac{ \delta_i+\delta_j}{2} \Theta \left( 1-\frac{r_{ij}}{\sigma_{ij}} \right), 
\end{equation}
for the repulsive linear spring potential, and the associated elastic modulus,
\begin{equation}
G_{\eta} = - \frac{d\Sigma_{\eta}}{d\eta}. 
\end{equation}
As discussed for applied simple shear strain, $G_{\eta}$ can also be decomposed into the affine and nonaffine contributions: $G_{\eta}=G^a_{\eta} + G^{na}_{\eta}$. For repulsive linear spring interactions, the affine contribution can be calculated analytically, 
which becomes
\begin{equation}
\label{eta_modulus}
G^a_{\eta}=\epsilon \sum^N_{i=1} \sum^N_{j>i} \left( \frac{\delta_i + \delta _j}{2} \right)^2 \frac{r_{ij}}{\sigma_{ij}^3} \left(3\frac{r_{ij}}{\sigma_{ij}} - 2 \right) \Theta \left( 1-\frac{r_{ij}}{\sigma_{ij}} \right).
\end{equation}
We measure $U$, $\Sigma_{\eta}$, $G_{\eta}$, and $G^a_{\eta}$ as a function of $\eta$ and identify jump and point changes in the 
interparticle contact network.

\subsection{Polydispersity strain at fixed pressure}
\label{poly_p}

To increase the polydispersity at fixed pressure, we take small steps in $\eta$ and minimize the enthalpy after each step until $\sum_{i=1}^N  | {\vec \nabla}_{ {\vec r}_i,L_x} H| < 10^{-13}$. During the applied strain, we measure the enthalpy, its derivative $dH/d\eta$, and changes in the interparticle contact network.

\subsection{Isotropic compression}

We also study the response of jammed packings to isotropic compression.  We compress the system by decreasing the box size in both dimensions by $-2 \Delta L/L =\Delta \phi/\phi$.  At the same time, we transform the particle coordinates by $x_i' = x^0_i \Delta L/L$ and $y_i' = y^0_i \Delta L/L$. After each compression step, we minimize the total potential energy until force balance is achieved. We measure the pressure,
\begin{equation}
p = -\frac{dU}{dA},
\end{equation}
which becomes
\begin{equation}
p=\frac{\epsilon}{2A} \sum^N_{i=1} \sum^N_{j>i} \left(1-\frac{r_{ij}}{\sigma_{ij}} \right) \frac{r_{ij}}{\sigma_{ij}} \Theta \left( 1-\frac{r_{ij}}{\sigma_{ij}} \right)
\end{equation}
for repulsive linear spring interactions, and the bulk modulus,
\begin{equation}
B = \phi \frac{dp}{d\phi}.
\end{equation}
$B$ can be decomposed into the affine and nonaffine contributions: $B=B^a + B^{na}$, respectively. For repulsive linear spring interactions, the affine contribution can be calculated analytically, 
\begin{equation}
\label{bulk_modulus}
B^a = \epsilon \frac{2\phi}{\pi N \langle \sigma^2 \rangle} \sum^N_{i=1} \sum^N_{j>i} \frac{r_{ij}^2}{\sigma_{ij}^2} \Theta \left( 1-\frac{r_{ij}}{\sigma_{ij}} \right).
\end{equation}
We calculate $B$ and $B^a$ as a function of packing fraction and monitor changes in the contact network during isotropic compression. 

\section{Results}
\label{results}

In this section, we present the results for the energy, stress, and elastic moduli for the five applied deformations described in Sec.~\ref{methods}.  We first show that changes in the interparticle contact networks during applied strain are either point changes or jump changes.  For a jump change, the  positions of the particles are discontinuous at the particular strain where the system becomes mechanically unstable and a particle rearrangement occurs.  In contrast, for a point change, an interparticle contact either breaks or a new contact forms as the particles move continuously during the applied strain.  We show that at a point change the derivative of the particle motions with respect to strain are discontinuous as are the derivatives of the potential energy/enthalpy, but at an order that depends on the interparticle potential. At small, but nonzero pressure, point changes occur in pairs over a range in strain. The first point change involves the formation of a new contact and the second involves the breaking of an existing contact. The difference in strain between these point changes decreases with pressure, and thus the pair of point changes coincide in the zero-pressure limit. To illustrate their importance, we detect exclusively point changes as we add polydispersity to originally monodisperse, ordered disk packings. Lastly, we present the statistics for jump and point changes for polydispersity strain applied at fixed packing fraction.

\subsection{Jump Changes}
\label{jump}

\begin{figure}[h!]
\centering
\includegraphics[height=7cm]{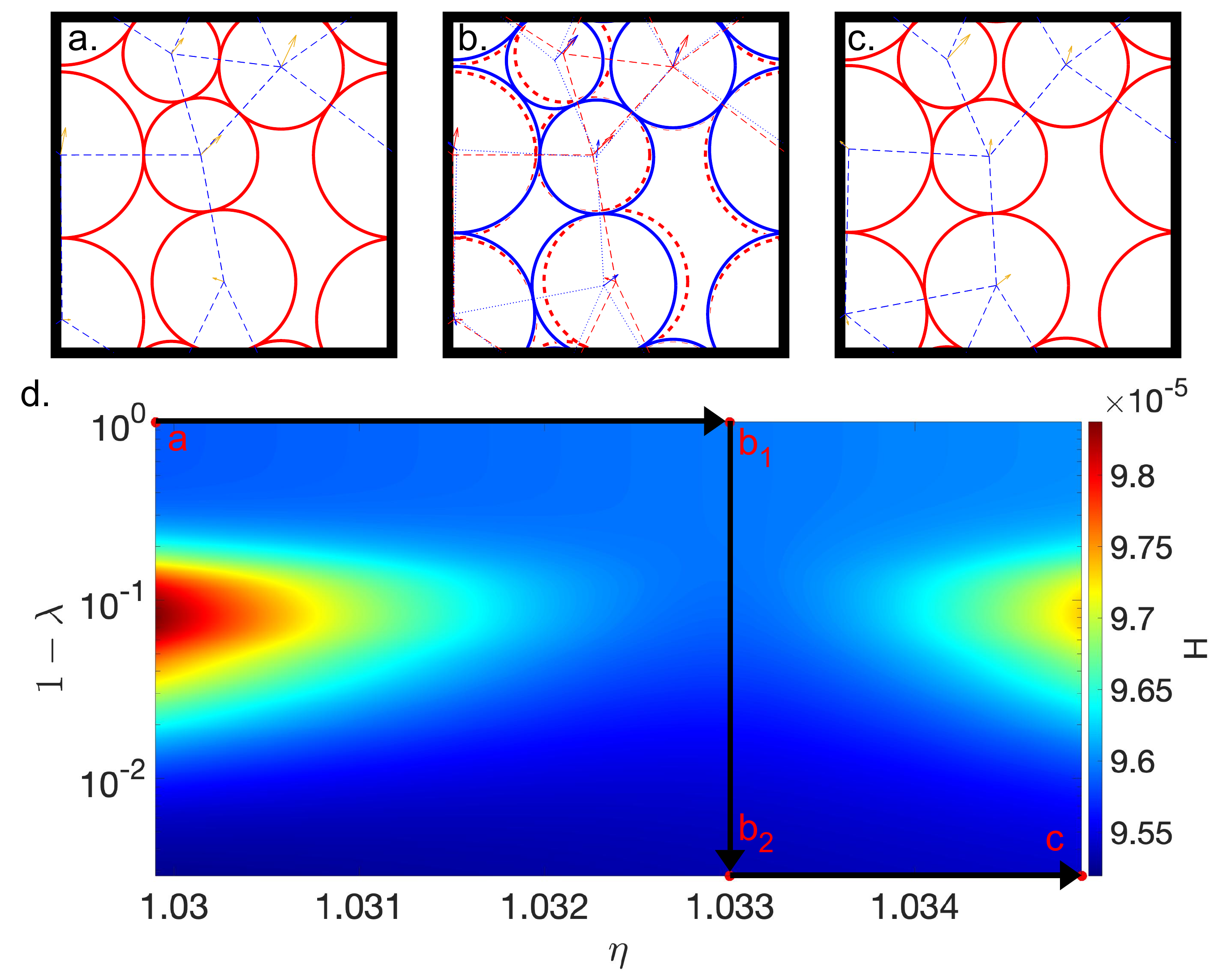}
\caption{An example of a jump change in a disordered packing of $N=6$ polydisperse disks during applied polydispersity strain at fixed pressure (Sec.~\ref{poly_p}). Panels (a)-(c) show the system, before, on both sides of, and after the change in the contact network. In (a) and (c), the red solid circles outline the particles, while the blue dashed lines represent the interparticle contact network. The arrows give the direction of the particle motion at the given value of strain. In (b), the blue solid circles (red dashed circles) represent the disk configuration and the blue dotted lines (red dashed lines) give the contact network after (before) the change. From the arrows and circles, we see that both the particle positions and directions of motion are discontinuous at the jump change. (d) The enthalpy $H$ (increasing from blue to red) is plotted as a function of the polydispersity $\eta$ and distance $\lambda$ along the path from the initial to the final state. The system starts in the upper left of the enthalpy landscape in the configuration in (a). At every $\eta$, the system can move vertically as long as the enthalpy decreases. The system is strained (increasing $\eta$) until it reaches point $b_1$, corresponding to the configuration in panel (b) with red dashed lines. After reaching $b_1$, a path to $b_2$ (the configuration in panel (b) with blue solid lines) opens and the system can reach a deeper local minimum without an increase in enthalpy during the trajectory. $\lambda$ parametrizes the distance along this path from $b_1$ to $b_2$. The system is then strained until point c, which corresponds to the configuration in panel (c). The bold black lines with arrows indicate the path taken.}
\label{fgr:Jump}
\end{figure}

We define a jump change as a change in the interparticle contact network for which the particle positions as a function of applied strain are discontinuous, i.e. the particles rearrange. The origin of the discontinuous particle motion stems from strain-induced changes in the energy or enthalpy landscape\cite{lacks,shear} and is illustrated in Fig.~\ref{fgr:Jump} for a disk packing undergoing polydispersity strain at fixed pressure (Sec.~\ref{poly_p}).  In Fig.~\ref{fgr:Jump} (a)-(c), we show the disk configurations before, during, and after a jump change. The enthalpy $H$ as a function of the polydispersity strain $\eta$ and the distance $\lambda$ along the path from the initial state before the jump change to the final state after the jump change is shown in Fig.~\ref{fgr:Jump} (d). To calculate $H(\eta,\lambda)$, we define a vector ${\vec \xi}=(L_x,x_1, \ldots, x_N, y_1, \ldots, y_N)$ that contains all of the degrees of freedom of the packing. If the path that the system takes from point $b_1$ to $b_2$ in Fig.~\ref{fgr:Jump} (d) is given by $\vec{\xi}(\eta^{*},\lambda)$, where the jump change occurs at $\eta^*$, $0 < \lambda < 1$, and $\Delta \vec{\xi} (\eta)=\vec{\xi}(\eta,1)-\vec{\xi}(\eta,0)$, then $\vec{\xi}(\eta,\lambda)=\Delta \vec{\xi} (\eta) ((\vec{\xi} (\eta^{*},\lambda)-\vec{\xi}(\eta^{*},0))/\Delta \vec{\xi} (\eta^*)$.  $\lambda$ parametrizes the path that the system takes in configuration space during enthalpy minimization from the initial state at $b_1$ ($\lambda=0$) to the final state at $b_2$ ($\lambda=1$). The system is strained by increasing $\eta$ in small steps (moving from left to right), followed by enthalpy minimization (moving vertically). The system begins in the upper left region of the landscape (point a), moves to the right (increasing $\eta$), and is initially prevented from moving toward the deeper minimum at the bottom of the enthalpy landscape by a barrier. As the system is further strained, the enthalpy barrier shrinks until the system reaches point $b_1$, where the barrier disappears, and the system evolves toward point $b_2$ with lower enthalpy. The disappearance of the enthalpy barrier at a given strain gives rise to the discontinuous change in the particle positions. We then continue straining the system until it reaches point c.   We find similar behavior for jump changes in the enthalpy landscape for systems undergoing simple shear strain at fixed pressure and in the energy landscape for systems undergoing simple shear strain or polydispersity strain at fixed packing fraction.
\subsection{Point Changes}
\label{point}
\begin{figure}[ht!]
\centering
\includegraphics[height=7cm]{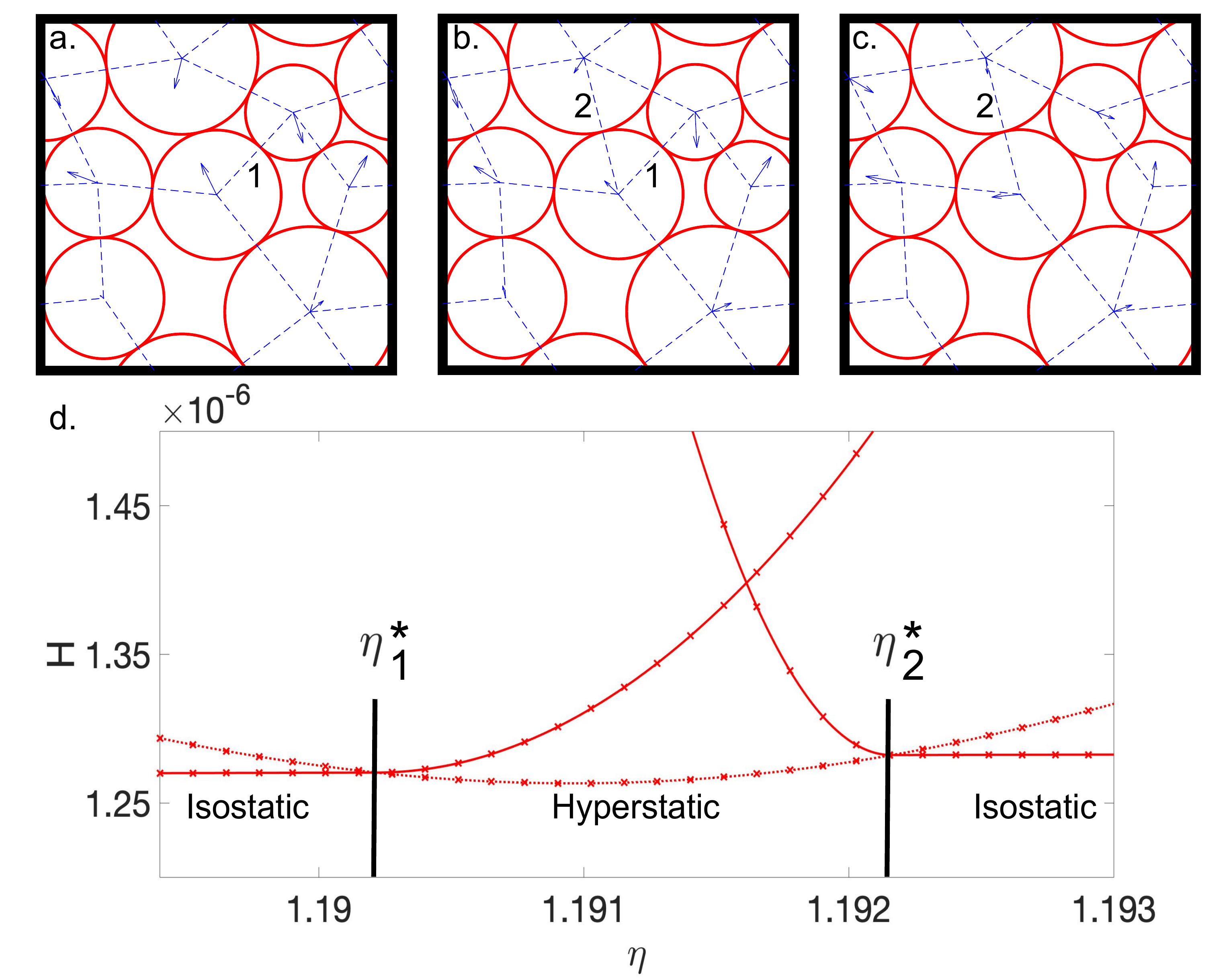}
\caption{An example of an $N=8$ polydisperse disk packing undergoing two successive point changes during applied polydispersity strain $\eta$ at fixed pressure. Panels (a) and (b) illustrate the first point change from an isostatic packing to a hyperstatic packing (with one extra contact) and (b) and (c) illustrate the second point change from the same hyperstatic packing to a different isostatic packing. All three packings are at target pressure $p_t=10^{-4}$. The arrows indicate the directions of particle motion at each strain. The number $1$ ($2$) labels the interparticle contact that is removed (added) during the two point changes. (d) Enthalpy $H$ plotted versus $\eta$ for the isostatic (hyperstatic) contact networks indicated by solid (dashed) lines. $\eta^*_1$ ($\eta^*_2$) labels the strain at which a contact is added (removed) from the contact network.}
\label{fgr:Point}
\end{figure}

We define a point change as the addition or removal of an interparticle contact at a given strain without discontinuous motion of the particles. The origin of a point change is that the positions of all particles for two or more distinct interparticle contact networks are the same at a given strain.  In Fig.~\ref{fgr:Point}, we illustrate two successive point changes for a disordered disk packing undergoing polydispersity strain at fixed pressure (Sec.~\ref{poly_p}).  In panels (a) and (b), we show the disk configurations corresponding to a point change from an isostatic packing to a hyperstatic packing with one additional contact, and in panels (b) and (c), we show the disk configurations corresponding to a point change from the same hyperstatic packing to a different isostatic packing. The arrows indicate the direction of motion of the particles, which show that the directions of the particle motion are not continuous over a point change. In Fig.~\ref{fgr:Point} (d), we show the enthlapy of the  configurations in panels (a)-(c) as a function of strain $\eta$ for target pressure $p_t=10^{-4}$.  We assume that (in the absence of changes in the contact network) the direction of the particle motion is constant with strain to extrapolate $H$ for the contact networks that are not enthalpy minima. 

At small $\eta$, the isostatic network in Fig.~\ref{fgr:Point} (a) has the lowest enthalpy of the three contact networks. At $1.190 < \eta^*_1 < 1.191$, $H$ of the configuration in (b) becomes less than that of the configuration in (a), and the system becomes hyperstatic with an additional interparticle contact.  At a higher strain $1.191 < \eta^{*}_2 < 1.192$, $H$ for the configuration in (c) becomes less than that of the configuration in (b), and the system transitions to a different isostatic contact network. Most importantly, the particle positions do not change discontinuously during each point change. In other words, the contact change happens between two energy minimized configurations. In contrast, for jump changes, as shown in Fig.~\ref{fgr:Jump} (d), the contact change occurs between a non-minimized configuration (point $b_1$) and a minimized configuration (point $b_2$). 

The changes of the particle trajectories in Fig.~\ref{fgr:Point} (a)-(c) demonstrate the importance of point changes. If contact $2$ did not form in panel (b), the two particles that form that contact would continue to move towards each other as they do in panel (a). These particle trajectories would cause a dramatic increase in enthalpy, as shown by $H(\eta)$ for the first isostatic contact network in panel (d). However, due to the formation of the new contact, the particle trajectories are altered following the point change as shown in panel (c). Despite the continuous particle motion that occurs during point changes, the particle trajectories are significantly altered with further strain.

\begin{figure}[h!]
\centering
\includegraphics[height=7cm]{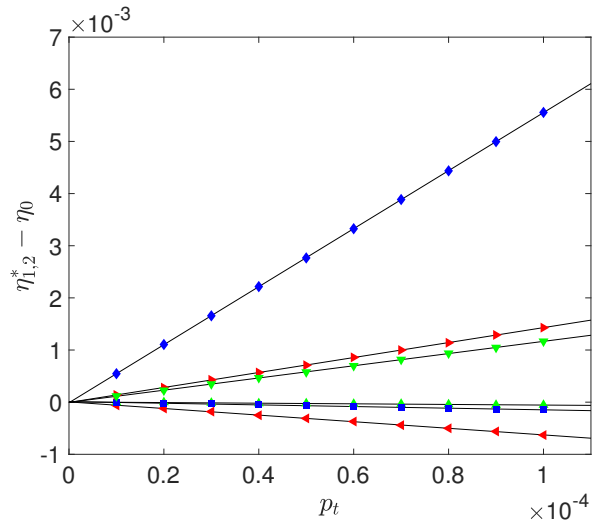}
\caption{For three sample polydisperse $N=8$ packings, we measure the polydispersity strain values at which the system transitions from an isostatic to a hyperstatic packing ($\eta^{*}_1$) and from the same hyperstatic packing to another isostatic packing ($\eta^{*}_2$) as shown in Fig.~\ref{fgr:Point}, at $10$ target pressures $p_t$.  We plot $\eta^{*}_{1,2}-\eta_0$, where $\eta_0 = \eta^*_{1,2}(p_t=0)$, versus $p_t$ for each contact change in each packing. The strain at which the packings transition from isostatic to hyperstatic, (i.e. between Fig.~\ref{fgr:Point} (a) and (b)), are represented by blue diamonds, red rightward triangles, and green downward triangles. The strains at which the packings transition from hyperstatic to isostatic, (i.e. between Fig.~\ref{fgr:Point} (b) and (c)) are represented by blue squares, red leftward triangles, and green upward triangles. Since all of the lines meet at $\eta^{*}_{1,2}=\eta_0$, the width of the hyperstatic strain region tends to zero in the $p_t=0$ limit.}
  \label{fgr:WidthHyperstatic}
\end{figure}

Fig.~\ref{fgr:WidthHyperstatic} displays the values of the polydispersity strain $\eta^*_1$ ($\eta_2^*$) at which several example polydisperse $N=8$ packings transition from an isostatic packing to a hyperstatic packing (and from the same hyperstatic packing to an isostatic packing) as a function of the target pressure $p_t$. For each packing, we find that both $\eta^*_1$ and $\eta^*_2$ are linear in $p_t$ with vertical intercept $\eta_0 = \eta^*_{1,2}(p_t=0)$. In Fig.~\ref{fgr:WidthHyperstatic}, we show that the values of $\eta^{*}_{1,2}$, corresponding to when the packing either gains a contact or loses a contact, possess the same $\eta_0$. Thus, the width of the strain region over which the system is hyperstatic between the two successive point changes (first from an isostatic packing to a hyperstatic packing and then from the same hyperstatic packing to another isostatic packing) tends to zero in the zero-pressure limit.  We find similar behavior for disk packings undergoing simple shear strain, as well as for larger system sizes. 

\subsection{Generalization of Jump and Point Changes to Other Strains}
\label{general}

While the illustrations of jump and point changes in Secs.~\ref{jump} and~\ref{point} considered polydispersity strain at constant pressure, all interparticle contact changes that occur during the applied strains that we consider (i.e. simple shear strain at constant packing fraction and at constant pressure, polydispersity strain at constant packing fraction and at constant pressure, and isotropic compression) can be classified as jump or point changes.  Further, we find that a point change at a given strain gives rise to continuous potential energy/enthalpy and its first derivatives, but causes discontinuities in the second derivatives of the potential energy/enthalpy at the given strain. The fact that the second derivatives of the potential energy/enthalpy are discontinuous (Eq.~\ref{shear_modulus}) is related to the repulsive linear spring interparticle potential that we employ; results for other finite-range repulsive potentials are discussed in Sec.~\ref{Hertzian}. In contrast, all jump changes give rise to discontinuities in the potential energy/enthalpy, as well as all of its derivatives, independent of the interparticle potential. 

\begin{figure}[h!]
\centering
\includegraphics[height=6cm]{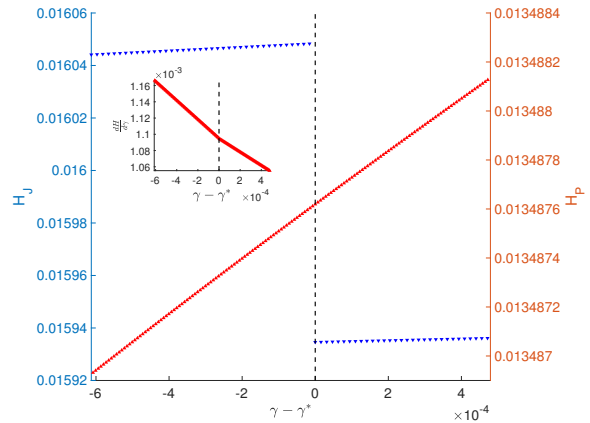}
\caption{The enthalpy $H_J$ (left vertical axis) and $H_P$ (right vertical axis) for a $N=16$ disk packing versus simple shear strain $\gamma-\gamma^*$ at constant pressure ($p=10^{-4}$) for a jump change (blue downward triangles) and a point change (red upward triangles) in the contact network at $\gamma^*$. For the jump change, there is a discontinuity in $H$ at $\gamma^*$. For the point change, both the enthalpy and its first derivative $dH/d\gamma$ (inset) are continuous at $\gamma^*$. However, the slope of $dH/d\gamma$ changes at $\gamma^*$, which indicates that $d^2 H/ d\gamma^2$ is discontinuous.}
\label{fgr:GammaPressure}
\end{figure}

As an example, in Fig.~\ref{fgr:GammaPressure}, we show the enthalpy $H$ as a function of simple shear strain $\gamma-\gamma^*$ for an $N=8$ packing (with repulsive linear spring interactions) undergoing simple shear at fixed pressure.  For the jump change at $\gamma^*$, $H$ is discontinuous. For the point change at $\gamma^*$, $H$ and $dH/d\gamma$ (in the inset) are both continuous, but $d^2 H/d\gamma^2$ is discontinuous.
The fact that the second derivative of the enthalpy, $G_{\gamma} + p_t (d^2 V / d\gamma^2)$, is discontinuous at a point change can be illustrated by analyzing the affine contribution of the shear modulus, $G^a_{\gamma}$ in Eq.~\ref{shear_modulus}, when contacts with zero overlap, $r_{ij} \rightarrow \sigma_{ij}$, are added to or removed from the contact network.  For the same reason, point changes give rise to discontinuities in the second derivatives with respect to strain of the potential energy/enthalpy for disk packings with repulsive linear spring interactions undergoing other applied strains. 

\subsection{Packing fraction-Strain Landscapes}
\label{landscape}

We refer to jammed disk packings with the same contact network as geometrical families (continuous regions) in the packing fraction and applied strain plane\cite{families2}. One can then consider contours of constant stress in the packing fraction and strain plane for each distinct contact network, and identify point and jump changes by calculating derivatives of the stress. In this section, we study the packing fraction and strain landscapes for both simple shear strain and polydispersity strain.  To construct these landscapes, we first generate a series of unjammed packings (with $\phi \approx 0.8$) over a range of strains. We find similar results using other packing fractions $\phi \lesssim \phi_J$. We then isotropically compress these packings (quasistatically) at each strain to packing fractions above jamming onset. For the disk packings at each packing fraction and strain, we measure the contact network, coordination number, and stress.  This protocol ensures that we can sample packings with both signs of the shear stress\cite{sheng}. For clarity, we show only a small portion of the strain-packing fraction landscape.

\begin{figure}[h!]
\centering
\includegraphics[height=7cm]{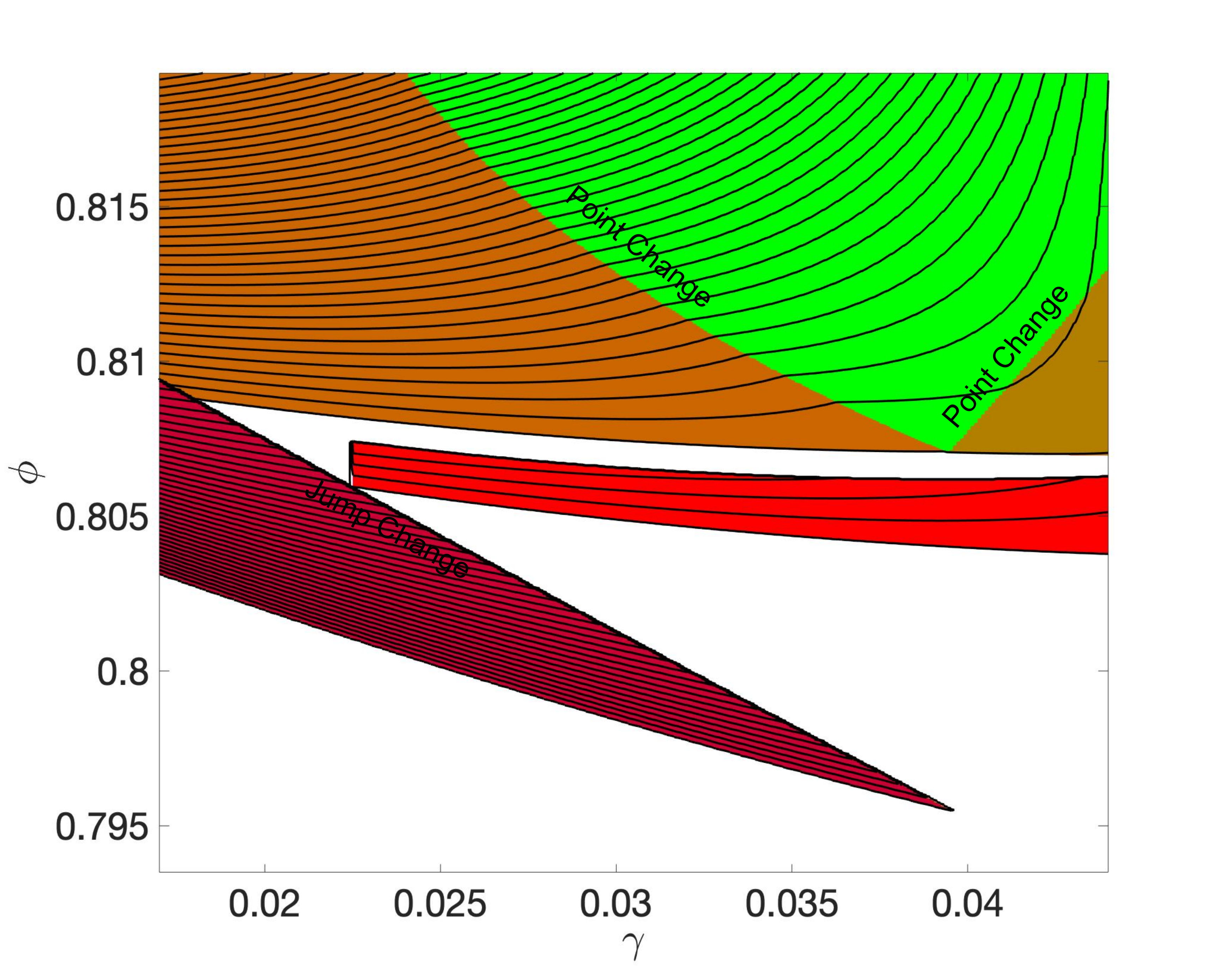}
\caption{An example landscape in the packing fraction $\phi$ and simple shear strain $\gamma$ plane for polydisperse $N=8$ disk packings. We first apply simple shear strain (quasistatically) at $\phi = 0.79$ (below jamming onset) to generate a series of unjammed packings over a range of $\gamma$ with step size $\Delta \gamma = 10^{-4}$. We then apply isotropic compression (quasistatically) with step size $\Delta \phi = 10^{-4}$ to these packings at each strain to packing fractions above jamming onset.  For each $\phi$ and $\gamma$, we show the contact network (color) and shear stress $\Sigma_{\gamma}$, where the lines are contours of constant $\Sigma_{\gamma}$ and the difference between adjacent lines is $\Delta \Sigma_{\gamma} \approx 7 \times 10^{-5}$. Jump changes can be identified by discontinuities in $\Sigma_{\gamma}$, and point changes by discontinuities in the derivative of $\Sigma_{\gamma}$. Red regions indicate isostatic contact networks, green regions indicate hyperstatic contact networks, and white regions indicate unjammed systems. Each region with a distinct red or green hue indicates packings with the same contact networks.}
\label{fgr:GammaPhiLandscape}
\end{figure}

In Fig.~\ref{fgr:GammaPhiLandscape}, we visualize polydisperse $N=8$ disk packings in the packing fraction $\phi$ and simple shear strain $\gamma$ plane. The color of a region indicates the type of contact network: regions that are red indicate isostatic contact networks and regions that are green indicate hyperstatic contact networks. Regions with different hues of red and green correspond to different contact networks. The white regions represent unjammed states. The lines provide contours of constant shear stress $\Sigma_{\gamma}$. $\Sigma_{\gamma}$ is discontinuous at jump changes, whereas it is continuous at point changes. 

The $\phi$-$\gamma$ landscape in Fig.~\ref{fgr:GammaPhiLandscape} has two lines of point changes, which can be traversed by compressing or decompressing the packing at fixed $\gamma$, by applying simple shear strain at fixed $\phi$, or by a combination of changes in $\phi$ and $\gamma$. The packing undergoes a point change when a contact is added (i.e. transitioning from an isostatic packing to a hyperstatic packing) or a contact is removed (i.e. transitioning from a hyperstatic packing to an isostatic packing). As discussed in Sec.~\ref{point}, the two lines of point changes merge into a single point near ($0.04$, $0.81$) in the zero-pressure limit. Traversing a point change in the forward direction leads to the same behavior as traversing it in the reverse direction. 

Lines of jump changes in Fig.~\ref{fgr:GammaPhiLandscape} occur when moving from an isostatic jammed region to an unjammed region. As we found for point changes, jump changes can be induced by compressing the packing at fixed $\gamma$, by applying simple shear strain at fixed $\phi$, or by a combination of changes in $\phi$ and $\gamma$. When undergoing a jump change to an unjammed state, the total potential energy and shear stress drop discontinuously from a finite value to zero. In Fig.~\ref{fgr:GammaPhiLandscape}, there is also a line of jump changes between two different isostatic packings near ($0.015$, $0.81$). 

Note that in Fig.~\ref{fgr:GammaPhiLandscape}, the system can transition from a jammed packing to unjammed packing through isotropic compression. Indeed, in recent computational studies, we showed that ``compression unjamming" occurs frequently near jamming onset.  We also showed that the probability for compression unjamming (averaged over a finite range of strain) approaches a finite value in the large-system limit, and thus compression unjamming occurs in the large-system limit\cite{kyle_prl}. 

\begin{figure}[h!]
\centering
\includegraphics[height=7cm]{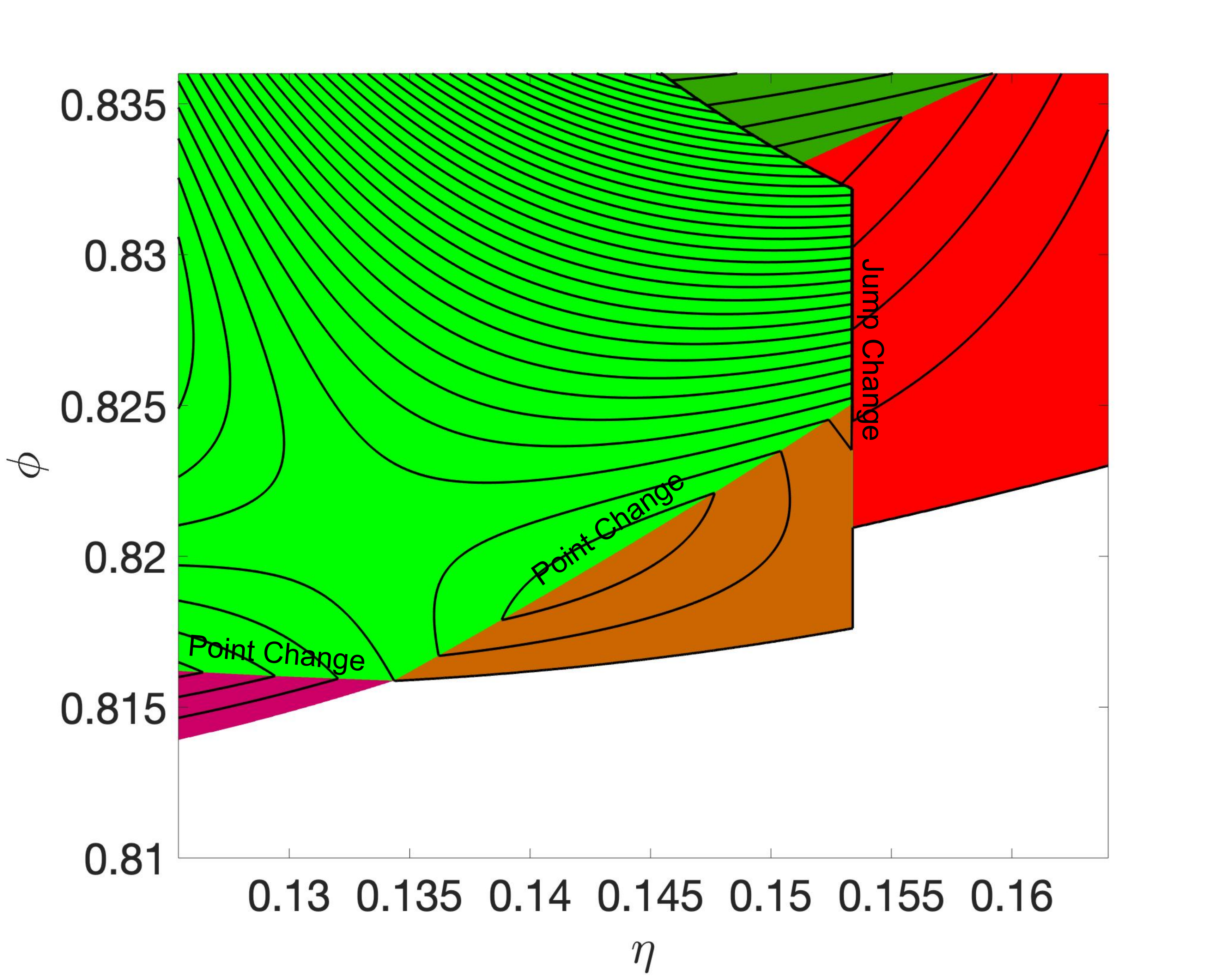}
\caption{An example landscape in the packing fraction $\phi$ and polydispersity strain $\eta$ plane for $N=8$ disk packings. We first apply polydispersity strain (quasistatically) at $\phi = 0.81$ (below jamming onset) to generate unjammed packings over a range of $\eta$ with step size $\Delta \eta=5\times 10^{-5}$. We then apply isotropic compression (with successive steps $\Delta \phi = 10^{-5}$ followed by energy minimization) to these packings at each strain to packing fractions above jamming onset. For each $\phi$ and $\eta$, we show the contact network (color) and stress $\Sigma_{\eta}$, where the lines are contours of constant $\Sigma_{\eta}$ and the difference between adjacent lines is $\Delta \Sigma_{\eta} \approx 2.5\times 10^{-4}$. Jump changes can be identified by discontinuities in $\Sigma_{\eta}$, and point changes by discontinuities in the derivative of $\Sigma_{\eta}$. Red regions indicate isostatic contact networks, green regions indicate hyperstatic contact networks, and white regions indicate unjammed systems. Each region with a distinct red or green hue indicates packings with the same contact networks.}
\label{fgr:EtaPhiLandscape}
\end{figure}

In Fig.~\ref{fgr:EtaPhiLandscape}, we show a portion of the packing fraction and polydispersity strain landscape for $N=8$ disk packings. The lines provide contours of constant polydispersity stress $\Sigma_{\eta}$. $\Sigma_{\eta}$ is discontinuous at jump changes, whereas it is continuous at point changes. In Fig.~\ref{fgr:EtaPhiLandscape}, there are two lines of point changes, which can be traversed by compressing or decompressing the packing at fixed $\eta$, by applying polydispersity strain at fixed $\phi$, or by a combination of changes in $\phi$ and $\eta$. Again, the two lines of point changes merge into a single point near ($0.135$, $0.815$) in the zero-pressure limit.  We find one line of jump changes in Fig.~\ref{fgr:EtaPhiLandscape} that can cause a transition between two isostatic packings, between a hyperstatic and an isostatic packing, and between two hyperstatic packings.

\subsection{Hertzian Spring Interactions}
\label{Hertzian}

\begin{figure}[h!]
\centering
\includegraphics[height=6cm]{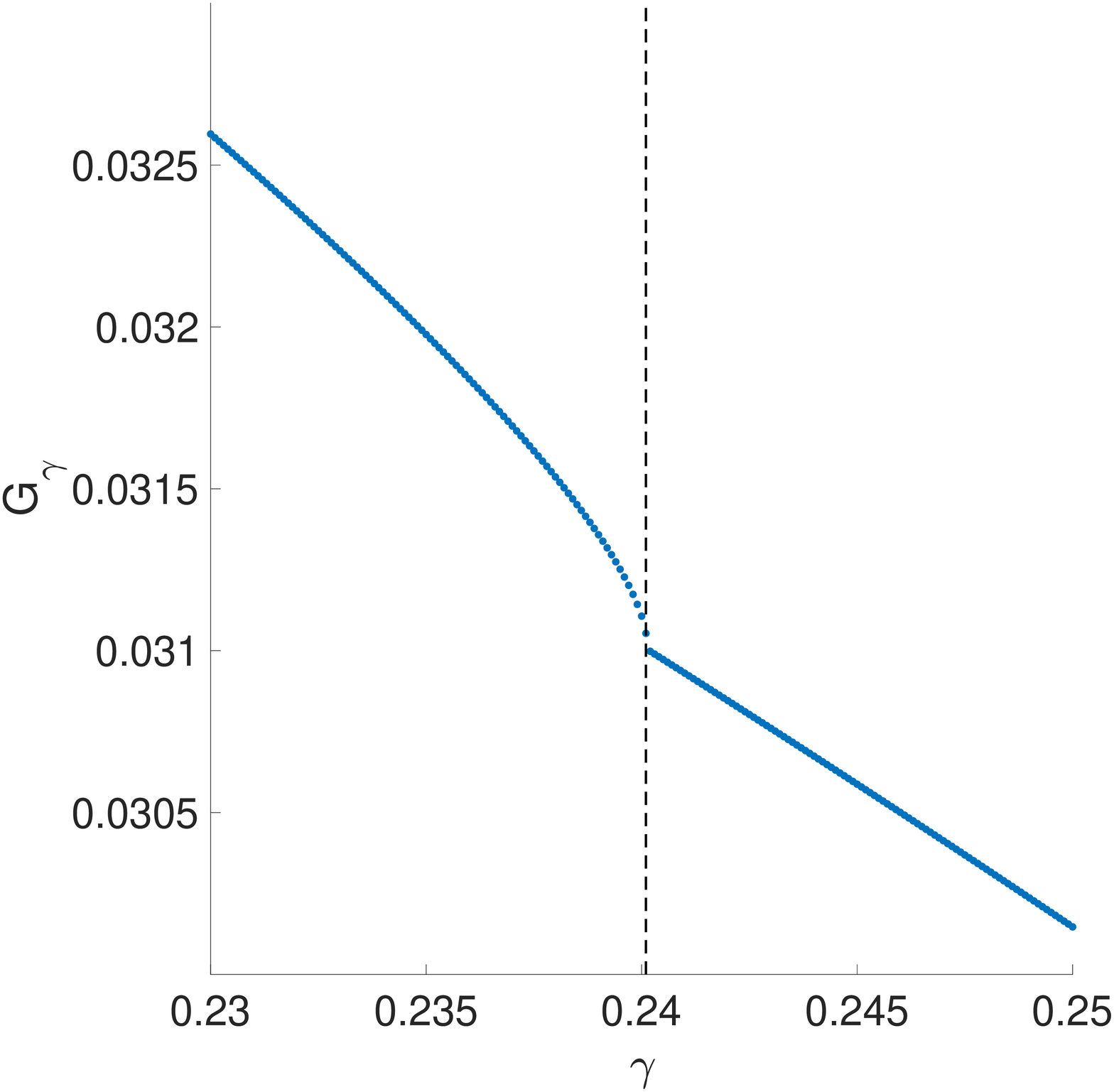}
\caption{Shear modulus $G_{\gamma}$ as a function of simple shear strain $\gamma$ (blue solid line) at fixed packing fraction $\phi=0.88$ for a $N=8$ disk packing with repulsive Hertzian spring interactions. The point change in the contact network (vertical dotted black line at $\gamma^* \approx 0.24$) does not cause a discontinuity in $G_{\gamma}$, but does cause a discontinuity in $dG_{\gamma}/d\gamma$.}
\label{fgr:HertzianG}
\end{figure}

In this section, we show preliminary results for frictionless disk packings that interact via repulsive Hertzian spring interactions ($\alpha = 5/2$ in Eq.~\ref{spring}) undergoing simple shear strain at fixed packing fraction. In Fig.~\ref{fgr:HertzianG}, we plot $G_{\gamma}$ versus $\gamma$ for a $N=16$ disk packing with repulsive Hertzian spring interactions across a point change. $G_{\gamma}$ is continuous across the point change, but $dG_{\gamma}/d\gamma$ is discontinuous.  This result can be anticipated by analyzing the affine contribution to the shear modulus for repulsive Hertzian spring interactions,
\begin{equation}
\label{hertz_mod}
\begin{aligned}
G^a_{\gamma} =&\epsilon \frac{L_x}{L_y^3} \sum_{i=1}^N \sum_{j>i}^N \sqrt{1-\frac{r_{ij}}{\sigma_{ij}}}\\
&\left( \frac{x_{ij}^2 y_{ij}^2}{\sigma_{ij} r_{ij}^3} \left(1+\frac{r_{ij}}{2\sigma_{ij}} \right) -\frac{y_{ij}^2}{\sigma_{ij} r_{ij}} \left(1-\frac{r_{ij}}{\sigma_{ij}} \right) \right).
\end{aligned}
\end{equation}
$G^a_{\gamma}$ for repulsive Hertzian spring interactions is similar to that for repulsive linear spring interactions (Eq.~\ref{shear_modulus}), but it has an additional factor of $\sqrt{1-r_{ij}/\sigma_{ij}}$. Thus, when a new contact is added to or removed from the contact network (at $r_{ij}=\sigma_{ij}$) during the applied strain, we expect that $G_{\gamma}$ will be continuous. If we take an additional derivative of $G^a_{\gamma}$ with respect to $\gamma$, the factor of $\sqrt{1-r_{ij}/\sigma_{ij}}$ moves to the denominator, and thus we expect that $dG_{\gamma}/d\gamma$ will be discontinuous across point changes, as shown in Fig.~\ref{fgr:HertzianG}. 

\subsection{Transition from a Hexagonal Crystal to a Disordered Crystal}
\label{disordered_crystal}

To illustrate the importance of point changes, we investigate the transition from a hexagonal crystal to a disordered crystal\cite{disordered_crystal1,disordered_crystal,disordered_crystal2} as a function of applied polydispersity strain in disk packings with repulsive linear spring interactions. The disordered crystal has properties in common with a hexagonal crystal (such as the disk positions and packing fraction), whereas other properties, such as the coordination number, stress, and elastic moduli, are similar to disordered, isostatic packings. Here, we show that the transition from the hexagonal crystal to the disordered crystal can be understood as series of point changes as a function of polydispersity strain, with no jump changes. We note that the transition to the disordered crystal can also be induced by simple shear and other applied strains.

\begin{figure}[h!]
\centering
\includegraphics[height=7cm]{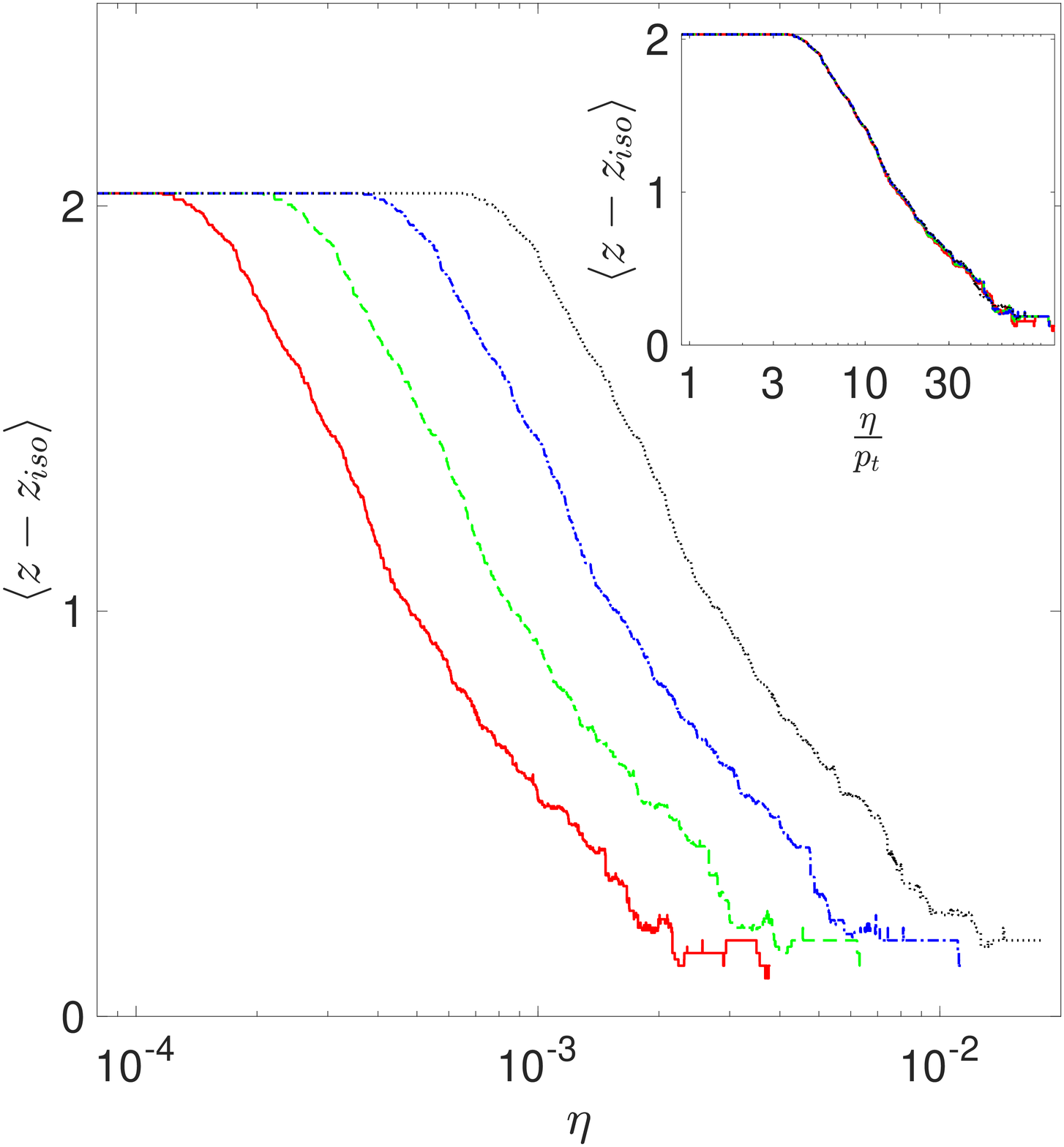}
\caption{The ensemble-averaged coordination number $\langle z-z_{\rm iso} \rangle$ versus the polydispersity strain $\eta$ at constant pressure $p_t$ for $N=64$ packings at $p_t =10^{-4.5}$ (red solid line), $10^{-4.25}$ (green dashed line), $10^{-4}$ (blue dot-dashed line), and $10^{-3.75}$ (black dotted line). The system was initialized in a hexagonal crystal at $\eta=0$. The inset shows that $\langle z-z_{\rm iso} \rangle$ can be collapsed by plotting it against $\eta/p_t$.  The data was averaged over $10$ packings for each pressure.}
\label{fgr:CrystalToIsoEnsemble}
\end{figure}

In Fig.~\ref{fgr:CrystalToIsoEnsemble}, we plot the ensemble-averaged excess coordination number $\langle z - z_{\rm iso} \rangle$, where $z_{\rm iso} = 2N_c^{\rm iso}/N$, as a function of polydispersity strain $\eta$ at fixed $p_t$.  $\langle z - z_{\rm iso} \rangle \approx 2$ at small $\eta$, and then begins to decrease toward zero at a characteristic $\eta_c$.  As shown in the inset to Fig.~\ref{fgr:CrystalToIsoEnsemble}, $\eta_c \sim p_t$ since $\langle z - z_{\rm iso} \rangle$ collapses when plotted versus $\eta/p_t$. Thus, in the zero-pressure limit, the hexagonal crystal at $\phi = \phi_x$ becomes isostatic with $z=z_{\rm iso}$  in the limit of zero applied strain. 

\begin{figure}[h!]
\centering
\includegraphics[height=6cm]{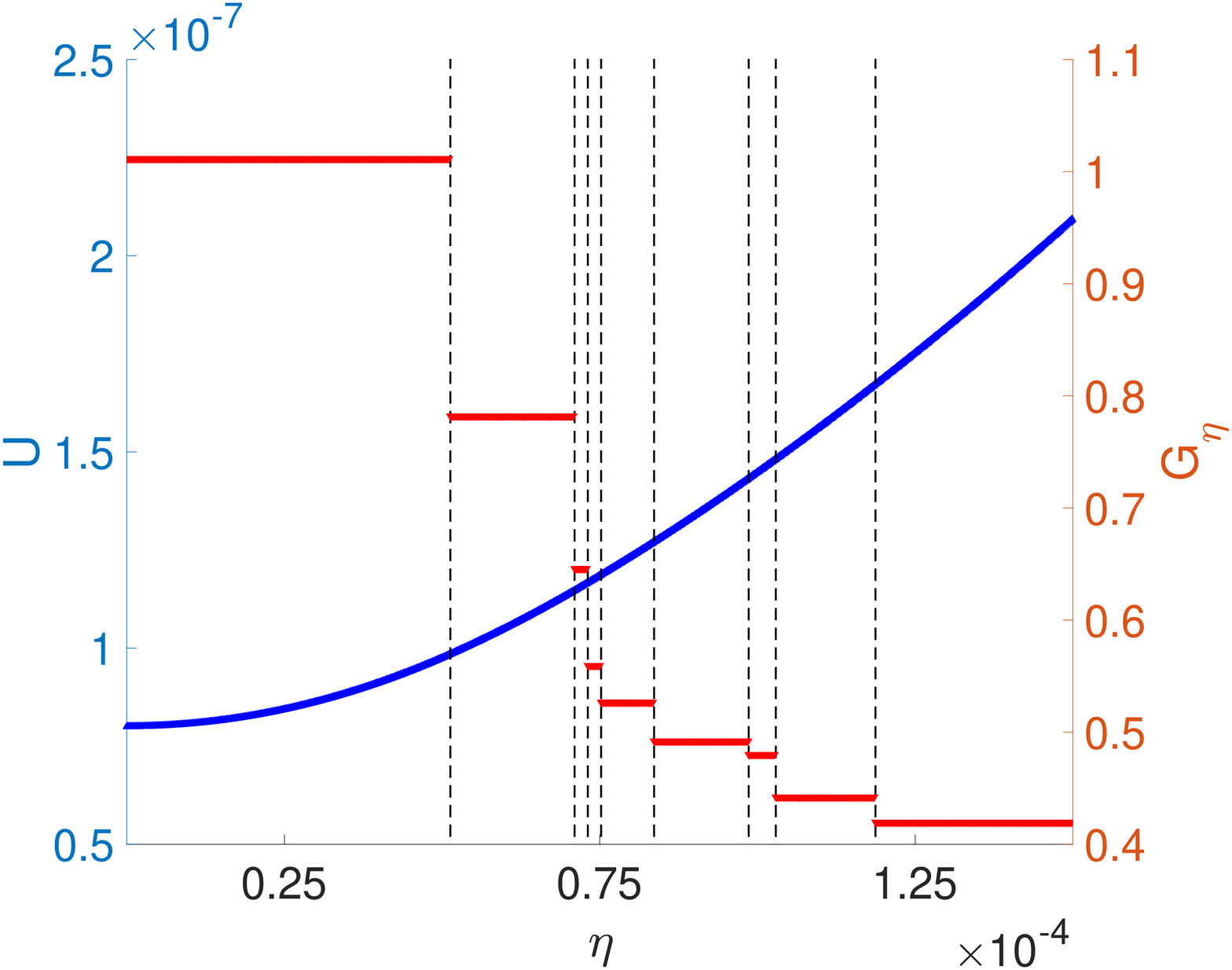}
\caption{The total potential energy $U$ (left vertical axis) and elastic modulus $G_{\eta}$ (right vertical axis) versus polydispersity strain $\eta$ at fixed packing fraction for a $N=16$ packing initialized in a hexagonal crystal at pressure $p=10^{-4}$. At each change in the contact network (black dashed vertical lines), $U$ (blue upward triangles) is continuous, while $G_{\eta}$ (red downward triangles) is discontinuous.}
\label{fgr:CrystalToIsoExample}
\end{figure}

\begin{figure}[h!]
\centering
\includegraphics[height=6cm]{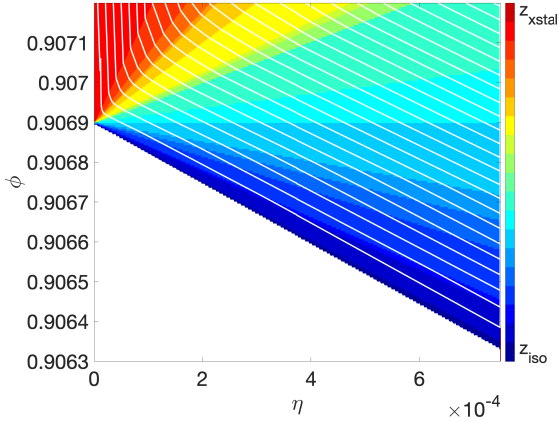}
\caption{The packing fraction $\phi$ and polydispersity strain $\eta$ landscape for a $N=16$ packing initialized in a hexagonal crystal. The color indicates the coordination number, ranging from isostatic with $z_{\rm iso} \sim 4$ to crystalline with $z=6$ (from blue to red). The white region corresponds to unjammed systems. The lines represent contours of constant polydispersity stress $\Sigma_{\eta}$ and the difference between adjacent lines is approximately $\Delta \Sigma_{\eta}=2\times 10^{-4}$. All of the changes in the contact networks are point changes, since there are no discontinuities in $\Sigma_{\eta}$.}
 \label{fgr:CrystalToIsoLandscape}
 \end{figure}

We find similar behavior for the transition from a hexagonal crystal to a disordered crystal when we apply polydispersity strain at fixed packing fraction.  In Fig.~\ref{fgr:CrystalToIsoExample}, we plot the total potential energy $U$ and elastic modulus $G_{\eta}$ versus $\eta$ at fixed $\phi$ for an $N=16$ packing initialized in a hexagonal crystal.  We show that at each change in the contact network $U$ is continuous, but $G_{\eta}$ is discontinuous, which signals that the changes in the contact network are point changes. In Fig.~\ref{fgr:CrystalToIsoLandscape}, we show the $\phi$-$\eta$ landscape for an $N=16$ packing initialized in a hexagonal crystal. There are many contact networks near the hexagaonal crystal, which are separated by point changes since there are no discontinuities in the polydispersity stress $\Sigma_{\eta}$.
In the zero-pressure limit, all of the point changes coincide and the system transitions from a hexagonal network to an isostatic network at zero strain.

\subsection{Distinguishing Point and Jump Contact Changes}
\label{Distinguishing}

\begin{figure}[h!]
\centering
\includegraphics[height=6cm]{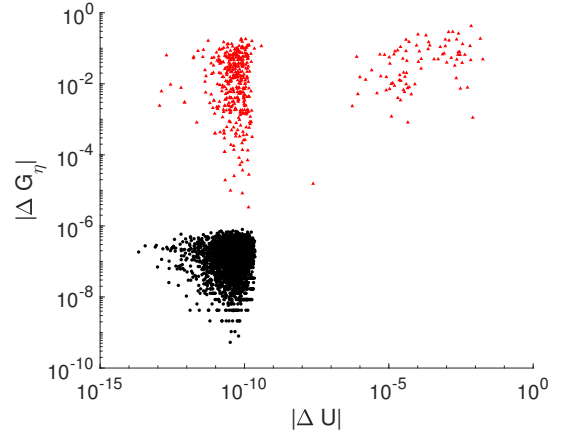}
\caption{A scatter plot of the absolute values of changes in the potential energy $|\Delta U|$ and shear modulus $|\Delta G_{\eta}|$ between successive polydispersity strain steps $\Delta \eta$ at constant packing fraction $\phi=0.88$ for $50$ $N=16$ packings. After every strain step, $U$ and $G_{\eta}$ were measured, and the difference between the values at the current step and the previous step was calculated to yield $\Delta U$ and $\Delta G_{\eta}$. The red triangles indicate a change in the contact network, whereas the black circles indicate strains where there was no change in the contact network.}
\label{fgr:EtaDisorderedStatistics}
\end{figure}

In this section, we discuss the changes in the total potential energy and elastic moduli that occur at point and jump changes for packings undergoing polydispersity strain at constant packing fraction.  In Fig.~\ref{fgr:EtaDisorderedStatistics}, we show a scatter plot of the absolute values of the changes in total potential energy $|\Delta U|$ and polydispersity modulus $|\Delta G_{\eta}|$ at polydispersity strains that correspond to changes in the contact network.  We also compare these values of $|\Delta U|$ and $|\Delta G_{\eta}|$ to those obtained from successive polydispersity strains where there is no change in the contact network. We find three distinct clusters of points: jump changes (with $|\Delta U|>10^{-9}$ and large values of $|\Delta G_{\eta}|$), point changes (with $|\Delta G_{\eta} |> 10^{-6}$ and small values of $|\Delta U|$), and points with small values of $|\Delta U|$ and $|\Delta G_{\eta}|$ where there are no changes in the contact network. This last set of points shifts to lower values of $|\Delta U|$ and $|\Delta G_{\eta}|$ with decreasing $\Delta \eta$ and improved force balance. All changes in the contact network during applied polydispersity strain can be classified as either point or jump changes. We find similar results for simple shear strain applied at fixed packing fraction and pressure, polydispersity strain applied at fixed pressure, and isotropic compression.

In principle, one can also use particle displacements (i.e. nonaffine particle motion) to identify changes in the contact networks\cite{meng}. For example, one could apply polydispersity strain from $\eta_1$ to $\eta_2$ yielding particle positions ${\vec r}(\eta_1)$ and ${\vec r}(\eta_2)$, and then reverse the strain from $\eta_2$ to $\eta_1$ to measure the new particle positions
${\vec r}'(\eta_1)$. The particle displacements $\Delta r = |{\vec r}(\eta_1) - {\vec r}'(\eta_1)|$ from this process will be large when there is a jump change between $\eta_1$ and $\eta_2$, whereas $\Delta r \rightarrow 0$ (in the small strain limit) for strain intervals where there is no jump change. Thus, measuring non-affine particle motions cannot be used to identify point changes. For this reason, we recommend measurements of $\Delta G$ and $\Delta U$ to identify point and jump changes in particulate media.

\section{Conclusions and future directions}
\label{conclusions}

In this article, we studied quasistatic deformations of jammed frictionless disk packings that interact via purely repulsive potentials as models of dense granular materials.  The deformations included simple shear strain at fixed packing fraction and at fixed pressure, polydispersity strain at fixed packing fraction and at fixed pressure, and isotropic compression.  We showed that there are two types of changes in the interparticle contact networks that occur during quasistatic deformation: point changes and jump changes. Jump changes involve changes in the contact network that are accompanied by discontinuous motion of the particles from one strain step to the next, whereas point changes involve small, continuous motion of the particles. It has been previously shown \cite{manning} that the relative frequency of these two types of events is constant with increasing system size.
Both types strongly affect the structural and mechanical properties of quasistatically deformed jammed granular systems.  For jump changes, the total potential energy (when the deformation is applied at constant packing fraction), or the enthalpy (when the deformation is applied at fixed pressure), as well as their derivatives with respect to strain are discontinuous. In contrast, point changes give rise to discontinuities in higher-order derivatives with respect to strain of the potential energy/enthalpy. For example, for disk packings with repulsive linear spring interactions, point changes cause discontinuities in the elastic moduli, which are proportional to second-order derivatives with respect to strain of the potential energy (when the deformation is applied at constant packing fraction) or the enthalpy (when the deformation is applied at constant pressure).  We then illustrated the important features of jump and point changes by showing contours of constant stress in the packing fraction and strain landscapes for the simple shear and polydispersity strain deformations. As a specific example of a physical phenomenon where point changes are dominant, we showed that the transition from a hexagonal crystal to a disordered crystal, which can possess an isostatic number of contacts, is caused by a series of point changes. 

The fact that point changes cause discontinuities with respect to strain in the second derivative of the potential energy/enthalpy (for disk packings with repulsive linear spring interactions) stems from the shape of the interparticle potential energy (Eq.~\ref{spring}).  The purely repulsive linear spring potential has a discontinuity in $d^2 U/dr_{ij}^2$ across a point change, and thus the elastic moduli, $G_{\gamma}$, $G_{\eta}$, and $B$, are discontinuous across a point change.  For the purely repulsive Hertzian spring potential with $\alpha = 5/2$ in Eq.~\ref{spring}, $d^3 U/dr_{ij}^3$ is discontinuous across a point change, and thus the derivatives of the elastic moduli with respect to strain (not the moduli themselves) are discontinuous.  The discontinuities caused by point changes will occur in higher-order derivatives of the potential energy (when the strain is applied at constant packing fraction) if higher-order derivatives of the interparticle potential are continuous. Similar results are found for the derivatives of the enthalpy when the strain is applied at fixed pressure. 

These results raise several important questions for future research. First, how do jammed packings behave when the applied strain is reversed\cite{dennin,regev,sastry} after point and jump changes occur in the interparticle contact networks? Point changes are completely reversible, since the particle motions are continuous during a point change.  Jump changes, however, are not reversible in this way. As shown in Fig.~\ref{fgr:Jump}, the packing immediately after the jump change has a lower potential energy (in the case of applied strain at constant packing fraction) than the packing immediately before the jump change.  Thus, when the strain is reversed after the jump change, the system will follow a different path in the energy landscape (than the one followed during the forward strain). However, it is possible that the system can undergo a series of point changes or another jump change during the reversed strain and return to the path in the energy landscape that was traversed during the forward strain. This behavior was termed ``loop reversibility" in Ref.~\cite{loop} and ``limit cycle" behavior in Ref.~\cite{cyclic_sand}, both of which studied systems undergoing cyclic simple shear strain.

In recent studies~\cite{kyle_prl}, we found that changes in the contact network during isotropic compression of jammed packings give rise to the power-law scaling of the shear modulus with pressure, i.e. $G_{\gamma} \sim p^{1/2}$ for repulsive linear spring interactions in $d=2$ and $3$.  Since both point and jump changes cause jumps in the shear modulus, $\Delta G_{\gamma}$, an interesting question is to determine whether point changes, jump changes, or both contribute significantly to the increase in the shear modulus during isotropic compression.  In addition, $G_{\gamma} \sim p^{2/3}$ for Hertzian spring interactions undergoing isotropic compression in $d=2$ and $3$\cite{jamming}. In future studies, we will investigate how jump and point changes give rise to this behavior, given that point changes do not cause discontinuities in $G_{\gamma}$ for Hertzian interactions. 

To understand the mechanical response of jammed packings to applied strain, one must be able to predict the potential energy (and other physical quantities that depend on the particle positions) as the system evolves along geometrical families, as well as across point and jump changes. We emphasize that it is still important to study point changes in packings undergoing quasistatic deformation even if the interparticle potential does not possess discontinuities in its derivatives. Even if there are no discontinuities in the interparticle potential, the particle trajectories change directions when the system undergoes each point change, which influences the evolution of the potential energy, stress, and elastic moduli as a function of strain. 

Another possible extension of the current studies is to investigate how point changes behave in packings of non-spherical particles. For example, in packings of circulo-lines in 2D, two particles with an ``end-end" contact behave differently than two particles with an ``end-middle" contact\cite{Hypostatic}. It will be interesting to study packings of circulo-lines that transition between these two types of contacts and determine whether this process can be described as a generalized point change, even though the interparticle contact network does not change.

A similar effect can occur in packings of spherical particles with frictional interactions. Numerous studies have shown that in addition to the number of contacts per particle, the ratio of the tangential to the normal force, $\zeta_{ij}$, at each contact between particles $i$ and $j$, plays an important role in determining the mechanical stability of frictional packings\cite{silbertfriction}. Thus, it is possible that effective ``point changes" can occur if $\zeta_{ij}$ varies significantly during strain even though particles $i$ and $j$ remain in contact.  

\section*{Conflicts of interest}
There are no conflicts to declare. 

\section*{Acknowledgements}
We acknowledge support from NSF Grant Nos. CBET-1605178
(K. V. and C. O.), DBI-1755494 (P. T.), CBET-2002782 (C. O. and J. Z.), and CBET-2002797 (M. S.), National Natural Science Foundation of China Grants No. 11572004 (Y. Y.) and
11972047 (Y. Y.), and the China Scholarship Council Grant Nos.
201806010289 (Y. Y.) and 201906340202 (S. Z.). This work was also supported by the High
Performance Computing facilities operated by Yale’s Center for
Research Computing and computing resources provided by the
Army Research Laboratory Defense University Research Instrumentation
Program Grant No. W911NF1810252.

\section*{Appendix I: Isotropic compression}
\label{appendix1}

\begin{figure}[h!]
\centering
\includegraphics[height=6cm]{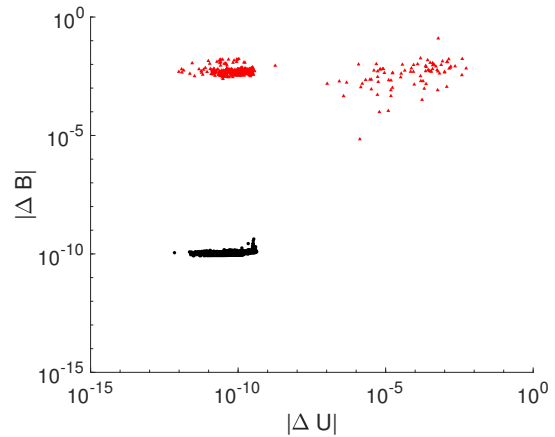}
\caption{A scatter plot of the absolute values of changes in the potential energy $|\Delta U|$ and bulk modulus $|\Delta B|$ between successive compression steps $\Delta \phi$ for $50$ $N=16$ packings. After every strain step, $U$ and $B$ were measured, and the difference between the values of the potential energy and bulk modulus at the current step and the previous step was calculated to yield $\Delta U$ and $\Delta B$. The red triangles indicate a change in the contact network, whereas the black circles indicate strains where there was no change in the contact network.}
\label{fgr:PhiDisorderedStatistics}
\end{figure}

In this Appendix, we show that the results for isotropic compression are similar to the results for the other strains that we studied. In Fig.~\ref{fgr:PhiDisorderedStatistics}, we show a scatter plot of the absolute values of the changes in total potential energy $|\Delta U|$ and bulk modulus $|\Delta B|$ at compression values that correspond to changes in the contact network.  We also compare these values of $|\Delta U|$ and $|\Delta B|$ to those obtained from successive compression steps where there is no change in the contact network. We find three distinct clusters of points: jump changes (with $|\Delta U|>10^{-7}$ and large values of $|\Delta B|$), point changes (with $|\Delta B|> 10^{-4}$ and small values of $|\Delta U|$), and points with small values of $|\Delta U|$ and $|\Delta B|$ where there are no changes in the contact network. This last set of points shifts to lower values of $|\Delta U|$ and $|\Delta B|$ with decreasing compression step size and improved force balance. (See Appendix II.) All changes in the contact network during applied compression can be classified as either point or jump changes. 

\section*{Appendix II: System Size Dependence}
\label{appendix2}

In this Appendix, we show that the presence of point and jump changes and our method for distinguishing between them do not change with increasing system size. For most of the results in this article, we used small systems with $N=6$ to $16$ disks with periodic boundary conditions in the $x$- and $y$-directions. Since point and jump changes have not been described before in the literature, the main goal of this article is to illustrate the theoretical foundations of point and jump contact changes, not to provide statistics of point and jump changes in the large-system limit. In previous studies, it has been shown that the length of geometrical families decreases strongly with increasing system size\cite{families}, and thus it makes sense to illustrate point and jump changes using small systems, where one can clearly see the beginning and end of each family. Further, the threshold required on force balance on each particle necessary to identify point and jump changes decreases toward zero with increasing system size, and thus it is much less computationally costly to study point and jump changes in small systems.

\begin{figure}[h!]
\centering
\includegraphics[height=6cm]{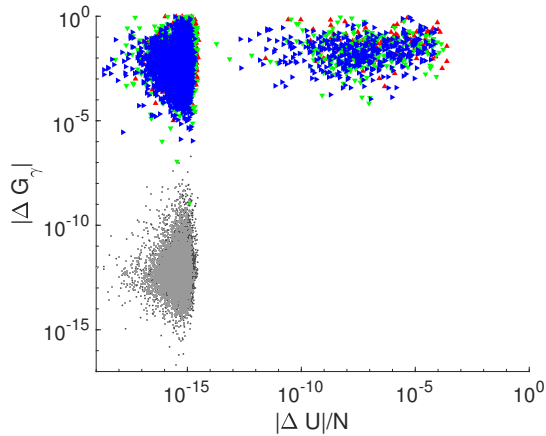}
\caption{A scatter plot of the absolute values of changes in the potential energy per particle $|\Delta U|/N$ and shear modulus $|\Delta G_{\gamma}|$ between successive shear steps $\Delta \gamma=7 \times 10^{-13}$ for $N=64$ (red upward triangles and black circles), $N=128$ (green downward triangles and dark gray dots), and $N=256$ (blue rightward triangles and light gray squares) packings. After every shear strain step, $U$ and $G_{\gamma}$ were measured, and the differences between the values at the current step and the previous step were calculated. The red, green, and blue triangles indicate a change in the contact network, whereas the black/gray points indicate strains where there was no change in the contact network.}
\label{fgr:SystemSize}
\end{figure}

Nevertheless, in Fig~\ref{fgr:SystemSize}, we show similar data as in Fig.~\ref{fgr:EtaDisorderedStatistics}, except for packings of $N=64$, $128$, and $256$ disks undergoing simple shear (with step size $\Delta \gamma = 7 \times 10^{-13}$) at fixed packing fraction $\phi=0.88$.  Again, we observe that there are three clusters of data points: one for jump changes (large $|\Delta U|/N$ and large $|\Delta G_{\gamma}|$), one for point changes (small $|\Delta U|/N$ and large $|\Delta G_{\gamma}|$), and one for the control group (small $|\Delta U|/N$ and small $|\Delta G_{\gamma}|$), for which point and jump changes do not occur. More importantly, we find that the location and spread of each of the three clusters remain the same for the three system sizes.

\begin{figure}[h!]
\centering
\includegraphics[height=6cm]{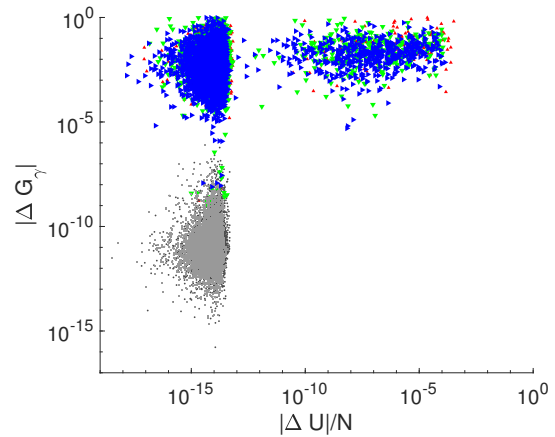}
\caption{A scatter plot of the absolute values of changes in the potential energy per particle $|\Delta U|/N$ and shear modulus $|\Delta G_{\gamma}|$ between successive shear steps $\Delta \gamma = 10^{-11}$ for $N=64$ (red upward triangles and black circles), $N=128$ (green downward triangles and dark gray dots), and $N=256$ (blue rightward triangles and light gray squares) packings. After every strain step, $U$ and $G_{\gamma}$ were measured, and the difference between the values at the current step and the previous step was calculated. The red, green, and blue triangles indicate a change in the contact network, whereas the black/gray points indicate strains where there was no change in the contact network.}
\label{fgr:AllBigStepSize}
\end{figure}

In Fig.~\ref{fgr:AllBigStepSize}, we show the same plot as in Fig.~\ref{fgr:SystemSize} for the three system sizes $N=64$, $128$, and $256$, except using a larger shear strain step size $\Delta \gamma = 10^{-11}$. The data points for $|\Delta G_{\gamma}|$ and $|\Delta U|/N$ corresponding to jump changes remain the same for the two shear strain step sizes. For the data points that correspond to point changes, the values of $|\Delta U|/N$ change with the shear strain step size, but the values of $|\Delta G_{\gamma}|$ do not.  In addition, for the points that do not correspond to changes in the contact network, both $|\Delta U|/N$ and $|\Delta G_{\gamma}|$ shift to larger values with the larger shear strain step size. Thus, $|\Delta U|/N \rightarrow 0$ and $|\Delta G_{\gamma}| \rightarrow 0$ in the limit $\Delta \gamma \rightarrow 0$ for data points that do not correspond to changes in the contact network. 

\nocite{*} 
\bibliography{Bibliography} 

\begin{thebibliography}{42}%
\makeatletter
\providecommand \@ifxundefined [1]{%
 \@ifx{#1\undefined}
}%
\providecommand \@ifnum [1]{%
 \ifnum #1\expandafter \@firstoftwo
 \else \expandafter \@secondoftwo
 \fi
}%
\providecommand \@ifx [1]{%
 \ifx #1\expandafter \@firstoftwo
 \else \expandafter \@secondoftwo
 \fi
}%
\providecommand \natexlab [1]{#1}%
\providecommand \enquote  [1]{``#1''}%
\providecommand \bibnamefont  [1]{#1}%
\providecommand \bibfnamefont [1]{#1}%
\providecommand \citenamefont [1]{#1}%
\providecommand \href@noop [0]{\@secondoftwo}%
\providecommand \href [0]{\begingroup \@sanitize@url \@href}%
\providecommand \@href[1]{\@@startlink{#1}\@@href}%
\providecommand \@@href[1]{\endgroup#1\@@endlink}%
\providecommand \@sanitize@url [0]{\catcode `\\12\catcode `\$12\catcode
  `\&12\catcode `\#12\catcode `\^12\catcode `\_12\catcode `\%12\relax}%
\providecommand \@@startlink[1]{}%
\providecommand \@@endlink[0]{}%
\providecommand \url  [0]{\begingroup\@sanitize@url \@url }%
\providecommand \@url [1]{\endgroup\@href {#1}{\urlprefix }}%
\providecommand \urlprefix  [0]{URL }%
\providecommand \Eprint [0]{\href }%
\providecommand \doibase [0]{http://dx.doi.org/}%
\providecommand \selectlanguage [0]{\@gobble}%
\providecommand \bibinfo  [0]{\@secondoftwo}%
\providecommand \bibfield  [0]{\@secondoftwo}%
\providecommand \translation [1]{[#1]}%
\providecommand \BibitemOpen [0]{}%
\providecommand \bibitemStop [0]{}%
\providecommand \bibitemNoStop [0]{.\EOS\space}%
\providecommand \EOS [0]{\spacefactor3000\relax}%
\providecommand \BibitemShut  [1]{\csname bibitem#1\endcsname}%
\let\auto@bib@innerbib\@empty
\bibitem [{\citenamefont {Behringer}\ and\ \citenamefont
  {Chakraborty}(2018)}]{jamming_review}%
  \BibitemOpen
  \bibfield  {author} {\bibinfo {author} {\bibfnamefont {R.~P.}\ \bibnamefont
  {Behringer}}\ and\ \bibinfo {author} {\bibfnamefont {B.}~\bibnamefont
  {Chakraborty}},\ }\href@noop {} {\bibfield  {journal} {\bibinfo  {journal}
  {Rep. on Prog. Phys.}\ }\textbf {\bibinfo {volume} {82}},\ \bibinfo {pages}
  {012601} (\bibinfo {year} {2018})}\BibitemShut {NoStop}%
\bibitem [{\citenamefont {Bi}\ \emph {et~al.}(2011)\citenamefont {Bi},
  \citenamefont {Zhang}, \citenamefont {Chakraborty},\ and\ \citenamefont
  {Behringer}}]{shear_jamming}%
  \BibitemOpen
  \bibfield  {author} {\bibinfo {author} {\bibfnamefont {D.}~\bibnamefont
  {Bi}}, \bibinfo {author} {\bibfnamefont {J.}~\bibnamefont {Zhang}}, \bibinfo
  {author} {\bibfnamefont {B.}~\bibnamefont {Chakraborty}}, \ and\ \bibinfo
  {author} {\bibfnamefont {R.~P.}\ \bibnamefont {Behringer}},\ }\href@noop {}
  {\bibfield  {journal} {\bibinfo  {journal} {Nature}\ }\textbf {\bibinfo
  {volume} {480}},\ \bibinfo {pages} {355} (\bibinfo {year}
  {2011})}\BibitemShut {NoStop}%
\bibitem [{\citenamefont {Bar\'{e}s}\ \emph {et~al.}(2017)\citenamefont
  {Bar\'{e}s}, \citenamefont {Wang}, \citenamefont {Want}, \citenamefont
  {Bertrand}, \citenamefont {O'Hern},\ and\ \citenamefont
  {Behringer}}]{avalanches}%
  \BibitemOpen
  \bibfield  {author} {\bibinfo {author} {\bibfnamefont {J.}~\bibnamefont
  {Bar\'{e}s}}, \bibinfo {author} {\bibfnamefont {D.}~\bibnamefont {Wang}},
  \bibinfo {author} {\bibfnamefont {D.}~\bibnamefont {Want}}, \bibinfo {author}
  {\bibfnamefont {T.}~\bibnamefont {Bertrand}}, \bibinfo {author}
  {\bibfnamefont {C.~S.}\ \bibnamefont {O'Hern}}, \ and\ \bibinfo {author}
  {\bibfnamefont {R.~P.}\ \bibnamefont {Behringer}},\ }\href@noop {} {\bibfield
   {journal} {\bibinfo  {journal} {Phys. Rev. E}\ }\textbf {\bibinfo {volume}
  {96}},\ \bibinfo {pages} {052902} (\bibinfo {year} {2017})}\BibitemShut
  {NoStop}%
\bibitem [{\citenamefont {Denisov}\ \emph {et~al.}(2016)\citenamefont
  {Denisov}, \citenamefont {L\"{o}rincz}, \citenamefont {Uhl}, \citenamefont
  {Dahmen},\ and\ \citenamefont {Schall}}]{avalanches2}%
  \BibitemOpen
  \bibfield  {author} {\bibinfo {author} {\bibfnamefont {D.~V.}\ \bibnamefont
  {Denisov}}, \bibinfo {author} {\bibfnamefont {K.~A.}\ \bibnamefont
  {L\"{o}rincz}}, \bibinfo {author} {\bibfnamefont {J.~T.}\ \bibnamefont
  {Uhl}}, \bibinfo {author} {\bibfnamefont {K.~A.}\ \bibnamefont {Dahmen}}, \
  and\ \bibinfo {author} {\bibfnamefont {P.}~\bibnamefont {Schall}},\
  }\href@noop {} {\bibfield  {journal} {\bibinfo  {journal} {Nature
  Communications}\ }\textbf {\bibinfo {volume} {7}},\ \bibinfo {pages} {010641}
  (\bibinfo {year} {2016})}\BibitemShut {NoStop}%
\bibitem [{\citenamefont {Mueth}\ \emph {et~al.}(2000)\citenamefont {Mueth},
  \citenamefont {Debregeas}, \citenamefont {Karczmar}, \citenamefont {Eng},
  \citenamefont {Nagel},\ and\ \citenamefont {Jaeger}}]{shear_banding1}%
  \BibitemOpen
  \bibfield  {author} {\bibinfo {author} {\bibfnamefont {D.~M.}\ \bibnamefont
  {Mueth}}, \bibinfo {author} {\bibfnamefont {G.~F.}\ \bibnamefont
  {Debregeas}}, \bibinfo {author} {\bibfnamefont {G.~S.}\ \bibnamefont
  {Karczmar}}, \bibinfo {author} {\bibfnamefont {P.~J.}\ \bibnamefont {Eng}},
  \bibinfo {author} {\bibfnamefont {S.~R.}\ \bibnamefont {Nagel}}, \ and\
  \bibinfo {author} {\bibfnamefont {H.~M.}\ \bibnamefont {Jaeger}},\
  }\href@noop {} {\bibfield  {journal} {\bibinfo  {journal} {Nature}\ }\textbf
  {\bibinfo {volume} {406}},\ \bibinfo {pages} {385} (\bibinfo {year}
  {2000})}\BibitemShut {NoStop}%
\bibitem [{\citenamefont {Karimi}\ and\ \citenamefont
  {Barrat}(2018)}]{shear_banding}%
  \BibitemOpen
  \bibfield  {author} {\bibinfo {author} {\bibfnamefont {K.}~\bibnamefont
  {Karimi}}\ and\ \bibinfo {author} {\bibfnamefont {J.-L.}\ \bibnamefont
  {Barrat}},\ }\href@noop {} {\bibfield  {journal} {\bibinfo  {journal}
  {Scientific Reports}\ }\textbf {\bibinfo {volume} {8}},\ \bibinfo {pages}
  {4021} (\bibinfo {year} {2018})}\BibitemShut {NoStop}%
\bibitem [{\citenamefont {Aranson}\ and\ \citenamefont
  {Tsimring}(2006)}]{collective}%
  \BibitemOpen
  \bibfield  {author} {\bibinfo {author} {\bibfnamefont {I.~S.}\ \bibnamefont
  {Aranson}}\ and\ \bibinfo {author} {\bibfnamefont {L.~S.}\ \bibnamefont
  {Tsimring}},\ }\href@noop {} {\bibfield  {journal} {\bibinfo  {journal} {Rev.
  Mod. Phys.}\ }\textbf {\bibinfo {volume} {78}},\ \bibinfo {pages} {641}
  (\bibinfo {year} {2006})}\BibitemShut {NoStop}%
\bibitem [{\citenamefont {O'Hern}\ \emph {et~al.}(2003)\citenamefont {O'Hern},
  \citenamefont {Silbert}, \citenamefont {Liu},\ and\ \citenamefont
  {Nagel}}]{jamming}%
  \BibitemOpen
  \bibfield  {author} {\bibinfo {author} {\bibfnamefont {C.~S.}\ \bibnamefont
  {O'Hern}}, \bibinfo {author} {\bibfnamefont {L.~E.}\ \bibnamefont {Silbert}},
  \bibinfo {author} {\bibfnamefont {A.~J.}\ \bibnamefont {Liu}}, \ and\
  \bibinfo {author} {\bibfnamefont {S.~R.}\ \bibnamefont {Nagel}},\ }\href@noop
  {} {\bibfield  {journal} {\bibinfo  {journal} {Phys. Rev. E}\ }\textbf
  {\bibinfo {volume} {68}},\ \bibinfo {pages} {011306} (\bibinfo {year}
  {2003})}\BibitemShut {NoStop}%
\bibitem [{\citenamefont {Liu}\ and\ \citenamefont {Nagel}(2010)}]{liu_review}%
  \BibitemOpen
  \bibfield  {author} {\bibinfo {author} {\bibfnamefont {A.~J.}\ \bibnamefont
  {Liu}}\ and\ \bibinfo {author} {\bibfnamefont {S.~R.}\ \bibnamefont
  {Nagel}},\ }\href@noop {} {\bibfield  {journal} {\bibinfo  {journal} {Ann.
  Rev. Condens. Matt. Phys.}\ }\textbf {\bibinfo {volume} {1}},\ \bibinfo
  {pages} {347} (\bibinfo {year} {2010})}\BibitemShut {NoStop}%
\bibitem [{\citenamefont {Tkachenko}\ and\ \citenamefont
  {Witten}(1999)}]{witten}%
  \BibitemOpen
  \bibfield  {author} {\bibinfo {author} {\bibfnamefont {A.~V.}\ \bibnamefont
  {Tkachenko}}\ and\ \bibinfo {author} {\bibfnamefont {T.~A.}\ \bibnamefont
  {Witten}},\ }\href@noop {} {\bibfield  {journal} {\bibinfo  {journal} {Phys.
  Rev. E}\ }\textbf {\bibinfo {volume} {60}},\ \bibinfo {pages} {687} (\bibinfo
  {year} {1999})}\BibitemShut {NoStop}%
\bibitem [{\citenamefont {Giacco}\ \emph {et~al.}(2017)\citenamefont {Giacco},
  \citenamefont {de~Arcangelis}, \citenamefont {Pica~Ciamarra},\ and\
  \citenamefont {Lippiello}}]{rattler}%
  \BibitemOpen
  \bibfield  {author} {\bibinfo {author} {\bibfnamefont {F.}~\bibnamefont
  {Giacco}}, \bibinfo {author} {\bibfnamefont {L.}~\bibnamefont
  {de~Arcangelis}}, \bibinfo {author} {\bibfnamefont {M.}~\bibnamefont
  {Pica~Ciamarra}}, \ and\ \bibinfo {author} {\bibfnamefont {E.}~\bibnamefont
  {Lippiello}},\ }\href@noop {} {\bibfield  {journal} {\bibinfo  {journal}
  {Soft Matter}\ }\textbf {\bibinfo {volume} {13}},\ \bibinfo {pages} {9132}
  (\bibinfo {year} {2017})}\BibitemShut {NoStop}%
\bibitem [{\citenamefont {Schreck}\ \emph {et~al.}(2011)\citenamefont
  {Schreck}, \citenamefont {O'Hern},\ and\ \citenamefont
  {Silbert}}]{hyperstatic}%
  \BibitemOpen
  \bibfield  {author} {\bibinfo {author} {\bibfnamefont {C.~F.}\ \bibnamefont
  {Schreck}}, \bibinfo {author} {\bibfnamefont {C.~S.}\ \bibnamefont {O'Hern}},
  \ and\ \bibinfo {author} {\bibfnamefont {L.}~\bibnamefont {Silbert}},\
  }\href@noop {} {\bibfield  {journal} {\bibinfo  {journal} {Phys. Rev. E}\
  }\textbf {\bibinfo {volume} {84}},\ \bibinfo {pages} {011305} (\bibinfo
  {year} {2011})}\BibitemShut {NoStop}%
\bibitem [{\citenamefont {Shen}\ \emph {et~al.}(2012)\citenamefont {Shen},
  \citenamefont {O'Hern},\ and\ \citenamefont {Shattuck}}]{shen}%
  \BibitemOpen
  \bibfield  {author} {\bibinfo {author} {\bibfnamefont {T.}~\bibnamefont
  {Shen}}, \bibinfo {author} {\bibfnamefont {C.~S.}\ \bibnamefont {O'Hern}}, \
  and\ \bibinfo {author} {\bibfnamefont {M.~D.}\ \bibnamefont {Shattuck}},\
  }\href@noop {} {\bibfield  {journal} {\bibinfo  {journal} {Phys. Rev. E}\
  }\textbf {\bibinfo {volume} {85}},\ \bibinfo {pages} {011308} (\bibinfo
  {year} {2012})}\BibitemShut {NoStop}%
\bibitem [{\citenamefont {Wyart}\ \emph
  {et~al.}(2005{\natexlab{a}})\citenamefont {Wyart}, \citenamefont {Nagel},\
  and\ \citenamefont {Witten}}]{wyart}%
  \BibitemOpen
  \bibfield  {author} {\bibinfo {author} {\bibfnamefont {M.}~\bibnamefont
  {Wyart}}, \bibinfo {author} {\bibfnamefont {S.~R.}\ \bibnamefont {Nagel}}, \
  and\ \bibinfo {author} {\bibfnamefont {T.~A.}\ \bibnamefont {Witten}},\
  }\href@noop {} {\bibfield  {journal} {\bibinfo  {journal} {Europhys. Lett.}\
  }\textbf {\bibinfo {volume} {72}},\ \bibinfo {pages} {486} (\bibinfo {year}
  {2005}{\natexlab{a}})}\BibitemShut {NoStop}%
\bibitem [{\citenamefont {Wyart}\ \emph
  {et~al.}(2005{\natexlab{b}})\citenamefont {Wyart}, \citenamefont {Silbert},
  \citenamefont {Nagel},\ and\ \citenamefont {Witten}}]{silbert}%
  \BibitemOpen
  \bibfield  {author} {\bibinfo {author} {\bibfnamefont {M.}~\bibnamefont
  {Wyart}}, \bibinfo {author} {\bibfnamefont {L.~E.}\ \bibnamefont {Silbert}},
  \bibinfo {author} {\bibfnamefont {S.~R.}\ \bibnamefont {Nagel}}, \ and\
  \bibinfo {author} {\bibfnamefont {T.~A.}\ \bibnamefont {Witten}},\
  }\href@noop {} {\bibfield  {journal} {\bibinfo  {journal} {Phys. Rev. E}\
  }\textbf {\bibinfo {volume} {72}},\ \bibinfo {pages} {051306} (\bibinfo
  {year} {2005}{\natexlab{b}})}\BibitemShut {NoStop}%
\bibitem [{\citenamefont {Goodrich}\ \emph {et~al.}(2012)\citenamefont
  {Goodrich}, \citenamefont {Liu},\ and\ \citenamefont {Nagel}}]{goodrich}%
  \BibitemOpen
  \bibfield  {author} {\bibinfo {author} {\bibfnamefont {C.~P.}\ \bibnamefont
  {Goodrich}}, \bibinfo {author} {\bibfnamefont {A.~J.}\ \bibnamefont {Liu}}, \
  and\ \bibinfo {author} {\bibfnamefont {S.~R.}\ \bibnamefont {Nagel}},\
  }\href@noop {} {\bibfield  {journal} {\bibinfo  {journal} {Phys. Rev. Lett.}\
  }\textbf {\bibinfo {volume} {109}},\ \bibinfo {pages} {095704} (\bibinfo
  {year} {2012})}\BibitemShut {NoStop}%
\bibitem [{\citenamefont {Chen}\ \emph {et~al.}(2018)\citenamefont {Chen},
  \citenamefont {Bertrand}, \citenamefont {Jin}, \citenamefont {Shattuck},\
  and\ \citenamefont {O'Hern}}]{sheng}%
  \BibitemOpen
  \bibfield  {author} {\bibinfo {author} {\bibfnamefont {S.}~\bibnamefont
  {Chen}}, \bibinfo {author} {\bibfnamefont {T.}~\bibnamefont {Bertrand}},
  \bibinfo {author} {\bibfnamefont {W.}~\bibnamefont {Jin}}, \bibinfo {author}
  {\bibfnamefont {M.~D.}\ \bibnamefont {Shattuck}}, \ and\ \bibinfo {author}
  {\bibfnamefont {C.~S.}\ \bibnamefont {O'Hern}},\ }\href@noop {} {\bibfield
  {journal} {\bibinfo  {journal} {Phys. Rev. E}\ }\textbf {\bibinfo {volume}
  {98}},\ \bibinfo {pages} {042906} (\bibinfo {year} {2018})}\BibitemShut
  {NoStop}%
\bibitem [{\citenamefont {Gao}\ \emph {et~al.}(2009)\citenamefont {Gao},
  \citenamefont {Blawzdziewicz},\ and\ \citenamefont {O'Hern}}]{families}%
  \BibitemOpen
  \bibfield  {author} {\bibinfo {author} {\bibfnamefont {G.-J.}\ \bibnamefont
  {Gao}}, \bibinfo {author} {\bibfnamefont {J.}~\bibnamefont {Blawzdziewicz}},
  \ and\ \bibinfo {author} {\bibfnamefont {C.~S.}\ \bibnamefont {O'Hern}},\
  }\href@noop {} {\bibfield  {journal} {\bibinfo  {journal} {Phys. Rev. E}\
  }\textbf {\bibinfo {volume} {80}},\ \bibinfo {pages} {061303} (\bibinfo
  {year} {2009})}\BibitemShut {NoStop}%
\bibitem [{\citenamefont {Bertrand}\ \emph {et~al.}(2016)\citenamefont
  {Bertrand}, \citenamefont {Behringer}, \citenamefont {Chakraborty},
  \citenamefont {O'Hern},\ and\ \citenamefont {Shattuck}}]{families2}%
  \BibitemOpen
  \bibfield  {author} {\bibinfo {author} {\bibfnamefont {T.}~\bibnamefont
  {Bertrand}}, \bibinfo {author} {\bibfnamefont {R.~P.}\ \bibnamefont
  {Behringer}}, \bibinfo {author} {\bibfnamefont {B.}~\bibnamefont
  {Chakraborty}}, \bibinfo {author} {\bibfnamefont {C.~S.}\ \bibnamefont
  {O'Hern}}, \ and\ \bibinfo {author} {\bibfnamefont {M.~D.}\ \bibnamefont
  {Shattuck}},\ }\href@noop {} {\bibfield  {journal} {\bibinfo  {journal}
  {Phys. Rev. E}\ }\textbf {\bibinfo {volume} {93}},\ \bibinfo {pages} {012901}
  (\bibinfo {year} {2016})}\BibitemShut {NoStop}%
\bibitem [{\citenamefont {VanderWerf}\ \emph {et~al.}(2020)\citenamefont
  {VanderWerf}, \citenamefont {Boromand}, \citenamefont {Shattuck},\ and\
  \citenamefont {O'Hern}}]{kyle_prl}%
  \BibitemOpen
  \bibfield  {author} {\bibinfo {author} {\bibfnamefont {K.}~\bibnamefont
  {VanderWerf}}, \bibinfo {author} {\bibfnamefont {A.}~\bibnamefont
  {Boromand}}, \bibinfo {author} {\bibfnamefont {M.~D.}\ \bibnamefont
  {Shattuck}}, \ and\ \bibinfo {author} {\bibfnamefont {C.~S.}\ \bibnamefont
  {O'Hern}},\ }\href@noop {} {\bibfield  {journal} {\bibinfo  {journal} {Phys.
  Rev. Lett.}\ }\textbf {\bibinfo {volume} {124}},\ \bibinfo {pages} {038004}
  (\bibinfo {year} {2020})}\BibitemShut {NoStop}%
\bibitem [{\citenamefont {Morse}\ \emph {et~al.}(2020)\citenamefont {Morse},
  \citenamefont {van Deen}, \citenamefont {Wijtmanns}, \citenamefont {van
  Heck},\ and\ \citenamefont {Manning}}]{manning}%
  \BibitemOpen
  \bibfield  {author} {\bibinfo {author} {\bibfnamefont {P.}~\bibnamefont
  {Morse}}, \bibinfo {author} {\bibfnamefont {M.}~\bibnamefont {van Deen}},
  \bibinfo {author} {\bibfnamefont {S.}~\bibnamefont {Wijtmanns}}, \bibinfo
  {author} {\bibfnamefont {M.}~\bibnamefont {van Heck}}, \ and\ \bibinfo
  {author} {\bibfnamefont {M.~L.}\ \bibnamefont {Manning}},\ }\href@noop {}
  {\bibfield  {journal} {\bibinfo  {journal} {Phys. Rev. Research}\ }\textbf
  {\bibinfo {volume} {2}},\ \bibinfo {pages} {023179} (\bibinfo {year}
  {2020})}\BibitemShut {NoStop}%
\bibitem [{\citenamefont {Malandro}\ and\ \citenamefont {Lacks}(1999)}]{lacks}%
  \BibitemOpen
  \bibfield  {author} {\bibinfo {author} {\bibfnamefont {D.~L.}\ \bibnamefont
  {Malandro}}\ and\ \bibinfo {author} {\bibfnamefont {D.~J.}\ \bibnamefont
  {Lacks}},\ }\href@noop {} {\bibfield  {journal} {\bibinfo  {journal} {J.
  Chem. Phys.}\ }\textbf {\bibinfo {volume} {110}},\ \bibinfo {pages} {4593}
  (\bibinfo {year} {1999})}\BibitemShut {NoStop}%
\bibitem [{\citenamefont {Cao}\ \emph {et~al.}(2018)\citenamefont {Cao},
  \citenamefont {Li}, \citenamefont {Kou}, \citenamefont {Xia}, \citenamefont
  {Li}, \citenamefont {Chen}, \citenamefont {Xie}, \citenamefont {Xiao},
  \citenamefont {Kob}, \citenamefont {Hong}, \citenamefont {Zhang},\ and\
  \citenamefont {Wang}}]{rearrangement}%
  \BibitemOpen
  \bibfield  {author} {\bibinfo {author} {\bibfnamefont {Y.}~\bibnamefont
  {Cao}}, \bibinfo {author} {\bibfnamefont {J.}~\bibnamefont {Li}}, \bibinfo
  {author} {\bibfnamefont {B.}~\bibnamefont {Kou}}, \bibinfo {author}
  {\bibfnamefont {C.}~\bibnamefont {Xia}}, \bibinfo {author} {\bibfnamefont
  {Z.}~\bibnamefont {Li}}, \bibinfo {author} {\bibfnamefont {R.}~\bibnamefont
  {Chen}}, \bibinfo {author} {\bibfnamefont {H.}~\bibnamefont {Xie}}, \bibinfo
  {author} {\bibfnamefont {T.}~\bibnamefont {Xiao}}, \bibinfo {author}
  {\bibfnamefont {W.}~\bibnamefont {Kob}}, \bibinfo {author} {\bibfnamefont
  {L.}~\bibnamefont {Hong}}, \bibinfo {author} {\bibfnamefont {J.}~\bibnamefont
  {Zhang}}, \ and\ \bibinfo {author} {\bibfnamefont {Y.}~\bibnamefont {Wang}},\
  }\href@noop {} {\bibfield  {journal} {\bibinfo  {journal} {Nature
  Communications}\ }\textbf {\bibinfo {volume} {9}},\ \bibinfo {pages} {2911}
  (\bibinfo {year} {2018})}\BibitemShut {NoStop}%
\bibitem [{\citenamefont {Mizuno}\ \emph
  {et~al.}(2016{\natexlab{a}})\citenamefont {Mizuno}, \citenamefont {Silbert},
  \citenamefont {Sperl}, \citenamefont {Mossa},\ and\ \citenamefont
  {Barrat}}]{cutoff}%
  \BibitemOpen
  \bibfield  {author} {\bibinfo {author} {\bibfnamefont {H.}~\bibnamefont
  {Mizuno}}, \bibinfo {author} {\bibfnamefont {L.}~\bibnamefont {Silbert}},
  \bibinfo {author} {\bibfnamefont {M.}~\bibnamefont {Sperl}}, \bibinfo
  {author} {\bibfnamefont {S.}~\bibnamefont {Mossa}}, \ and\ \bibinfo {author}
  {\bibfnamefont {J.-L.}\ \bibnamefont {Barrat}},\ }\href@noop {} {\bibfield
  {journal} {\bibinfo  {journal} {Phys. Rev. E}\ }\textbf {\bibinfo {volume}
  {93}},\ \bibinfo {pages} {043314} (\bibinfo {year}
  {2016}{\natexlab{a}})}\BibitemShut {NoStop}%
\bibitem [{\citenamefont {Goodrich}\ \emph {et~al.}(2014)\citenamefont
  {Goodrich}, \citenamefont {Liu},\ and\ \citenamefont
  {Nagel}}]{disordered_crystal1}%
  \BibitemOpen
  \bibfield  {author} {\bibinfo {author} {\bibfnamefont {C.~P.}\ \bibnamefont
  {Goodrich}}, \bibinfo {author} {\bibfnamefont {A.~J.}\ \bibnamefont {Liu}}, \
  and\ \bibinfo {author} {\bibfnamefont {S.~R.}\ \bibnamefont {Nagel}},\
  }\href@noop {} {\bibfield  {journal} {\bibinfo  {journal} {Nat. Phys.}\
  }\textbf {\bibinfo {volume} {10}},\ \bibinfo {pages} {578} (\bibinfo {year}
  {2014})}\BibitemShut {NoStop}%
\bibitem [{\citenamefont {Tong}\ \emph {et~al.}(2015)\citenamefont {Tong},
  \citenamefont {Tan},\ and\ \citenamefont {Xu}}]{disordered_crystal}%
  \BibitemOpen
  \bibfield  {author} {\bibinfo {author} {\bibfnamefont {H.}~\bibnamefont
  {Tong}}, \bibinfo {author} {\bibfnamefont {P.}~\bibnamefont {Tan}}, \ and\
  \bibinfo {author} {\bibfnamefont {N.}~\bibnamefont {Xu}},\ }\href@noop {}
  {\bibfield  {journal} {\bibinfo  {journal} {Scientific Reports}\ }\textbf
  {\bibinfo {volume} {5}},\ \bibinfo {pages} {15378} (\bibinfo {year}
  {2015})}\BibitemShut {NoStop}%
\bibitem [{\citenamefont {Schreck}\ \emph {et~al.}(2014)\citenamefont
  {Schreck}, \citenamefont {O'Hern},\ and\ \citenamefont
  {Shattuck}}]{hertzian}%
  \BibitemOpen
  \bibfield  {author} {\bibinfo {author} {\bibfnamefont {C.~F.}\ \bibnamefont
  {Schreck}}, \bibinfo {author} {\bibfnamefont {C.~S.}\ \bibnamefont {O'Hern}},
  \ and\ \bibinfo {author} {\bibfnamefont {M.~D.}\ \bibnamefont {Shattuck}},\
  }\href@noop {} {\bibfield  {journal} {\bibinfo  {journal} {Granular Matter}\
  }\textbf {\bibinfo {volume} {16}},\ \bibinfo {pages} {209} (\bibinfo {year}
  {2014})}\BibitemShut {NoStop}%
\bibitem [{\citenamefont {Arceri}\ and\ \citenamefont
  {Corwin}(2020)}]{unified}%
  \BibitemOpen
  \bibfield  {author} {\bibinfo {author} {\bibfnamefont {F.}~\bibnamefont
  {Arceri}}\ and\ \bibinfo {author} {\bibfnamefont {E.~I.}\ \bibnamefont
  {Corwin}},\ }\href@noop {} {\bibfield  {journal} {\bibinfo  {journal} {Phys.
  Rev. Lett.}\ }\textbf {\bibinfo {volume} {124}},\ \bibinfo {pages} {238002}
  (\bibinfo {year} {2020})}\BibitemShut {NoStop}%
\bibitem [{\citenamefont {Johnson}(1985)}]{johnson}%
  \BibitemOpen
  \bibfield  {author} {\bibinfo {author} {\bibfnamefont {K.~L.}\ \bibnamefont
  {Johnson}},\ }\href@noop {} {\emph {\bibinfo {title} {Contact Mechanics}}}\
  (\bibinfo  {publisher} {Cambridge University Press},\ \bibinfo {year}
  {1985})\BibitemShut {NoStop}%
\bibitem [{\citenamefont {Bitzek}\ \emph {et~al.}(2006)\citenamefont {Bitzek},
  \citenamefont {Koskinen}, \citenamefont {G\"{a}hler}, \citenamefont
  {Moseler},\ and\ \citenamefont {Gumbsch}}]{FIRE}%
  \BibitemOpen
  \bibfield  {author} {\bibinfo {author} {\bibfnamefont {E.}~\bibnamefont
  {Bitzek}}, \bibinfo {author} {\bibfnamefont {P.}~\bibnamefont {Koskinen}},
  \bibinfo {author} {\bibfnamefont {F.}~\bibnamefont {G\"{a}hler}}, \bibinfo
  {author} {\bibfnamefont {M.}~\bibnamefont {Moseler}}, \ and\ \bibinfo
  {author} {\bibfnamefont {P.}~\bibnamefont {Gumbsch}},\ }\href@noop {}
  {\bibfield  {journal} {\bibinfo  {journal} {Phys. Rev. Lett}\ }\textbf
  {\bibinfo {volume} {97}},\ \bibinfo {pages} {170201} (\bibinfo {year}
  {2006})}\BibitemShut {NoStop}%
\bibitem [{\citenamefont {Maloney}\ and\ \citenamefont
  {Lema\^{i}tre}(2006)}]{Lemaitre}%
  \BibitemOpen
  \bibfield  {author} {\bibinfo {author} {\bibfnamefont {C.}~\bibnamefont
  {Maloney}}\ and\ \bibinfo {author} {\bibfnamefont {A.}~\bibnamefont
  {Lema\^{i}tre}},\ }\href@noop {} {\bibfield  {journal} {\bibinfo  {journal}
  {Phys. Rev. E}\ }\textbf {\bibinfo {volume} {74}},\ \bibinfo {pages} {016118}
  (\bibinfo {year} {2006})}\BibitemShut {NoStop}%
\bibitem [{\citenamefont {Mizuno}\ \emph
  {et~al.}(2016{\natexlab{b}})\citenamefont {Mizuno}, \citenamefont {Saitoh},\
  and\ \citenamefont {Silbert}}]{mizuno}%
  \BibitemOpen
  \bibfield  {author} {\bibinfo {author} {\bibfnamefont {H.}~\bibnamefont
  {Mizuno}}, \bibinfo {author} {\bibfnamefont {K.}~\bibnamefont {Saitoh}}, \
  and\ \bibinfo {author} {\bibfnamefont {L.~E.}\ \bibnamefont {Silbert}},\
  }\href@noop {} {\bibfield  {journal} {\bibinfo  {journal} {Phys. Rev. E}\
  }\textbf {\bibinfo {volume} {93}},\ \bibinfo {pages} {062905} (\bibinfo
  {year} {2016}{\natexlab{b}})}\BibitemShut {NoStop}%
\bibitem [{\citenamefont {Blank-Burian}\ and\ \citenamefont
  {Heuer}(2018)}]{shear}%
  \BibitemOpen
  \bibfield  {author} {\bibinfo {author} {\bibfnamefont {M.}~\bibnamefont
  {Blank-Burian}}\ and\ \bibinfo {author} {\bibfnamefont {A.}~\bibnamefont
  {Heuer}},\ }\href@noop {} {\bibfield  {journal} {\bibinfo  {journal} {Phys.
  Rev. E}\ }\textbf {\bibinfo {volume} {98}},\ \bibinfo {pages} {033002}
  (\bibinfo {year} {2018})}\BibitemShut {NoStop}%
\bibitem [{\citenamefont {Acharya}\ \emph {et~al.}(2020)\citenamefont
  {Acharya}, \citenamefont {Sengupta}, \citenamefont {Chakraborty},\ and\
  \citenamefont {Ramola}}]{disordered_crystal2}%
  \BibitemOpen
  \bibfield  {author} {\bibinfo {author} {\bibfnamefont {P.}~\bibnamefont
  {Acharya}}, \bibinfo {author} {\bibfnamefont {S.}~\bibnamefont {Sengupta}},
  \bibinfo {author} {\bibfnamefont {B.}~\bibnamefont {Chakraborty}}, \ and\
  \bibinfo {author} {\bibfnamefont {K.}~\bibnamefont {Ramola}},\ }\href@noop {}
  {\bibfield  {journal} {\bibinfo  {journal} {Phys. Rev. Lett.}\ }\textbf
  {\bibinfo {volume} {124}},\ \bibinfo {pages} {168004} (\bibinfo {year}
  {2020})}\BibitemShut {NoStop}%
\bibitem [{\citenamefont {Fan}\ \emph {et~al.}(2017)\citenamefont {Fan},
  \citenamefont {Zhang}, \citenamefont {Schroers}, \citenamefont {Shattuck},\
  and\ \citenamefont {O'Hern}}]{meng}%
  \BibitemOpen
  \bibfield  {author} {\bibinfo {author} {\bibfnamefont {M.}~\bibnamefont
  {Fan}}, \bibinfo {author} {\bibfnamefont {K.}~\bibnamefont {Zhang}}, \bibinfo
  {author} {\bibfnamefont {J.}~\bibnamefont {Schroers}}, \bibinfo {author}
  {\bibfnamefont {M.~D.}\ \bibnamefont {Shattuck}}, \ and\ \bibinfo {author}
  {\bibfnamefont {C.~S.}\ \bibnamefont {O'Hern}},\ }\href@noop {} {\bibfield
  {journal} {\bibinfo  {journal} {Phys. Rev. E}\ }\textbf {\bibinfo {volume}
  {96}},\ \bibinfo {pages} {032602} (\bibinfo {year} {2017})}\BibitemShut
  {NoStop}%
\bibitem [{\citenamefont {Lundberg}\ \emph {et~al.}(2008)\citenamefont
  {Lundberg}, \citenamefont {Krishan}, \citenamefont {Xu}, \citenamefont
  {O'Hern},\ and\ \citenamefont {Dennin}}]{dennin}%
  \BibitemOpen
  \bibfield  {author} {\bibinfo {author} {\bibfnamefont {M.}~\bibnamefont
  {Lundberg}}, \bibinfo {author} {\bibfnamefont {K.}~\bibnamefont {Krishan}},
  \bibinfo {author} {\bibfnamefont {N.}~\bibnamefont {Xu}}, \bibinfo {author}
  {\bibfnamefont {C.~S.}\ \bibnamefont {O'Hern}}, \ and\ \bibinfo {author}
  {\bibfnamefont {M.}~\bibnamefont {Dennin}},\ }\href@noop {} {\bibfield
  {journal} {\bibinfo  {journal} {Phys. Rev. E}\ }\textbf {\bibinfo {volume}
  {77}},\ \bibinfo {pages} {041505} (\bibinfo {year} {2008})}\BibitemShut
  {NoStop}%
\bibitem [{\citenamefont {Regev}\ \emph {et~al.}(2015)\citenamefont {Regev},
  \citenamefont {Weber}, \citenamefont {Reichhardt}, \citenamefont {Dahmen},\
  and\ \citenamefont {Lookman}}]{regev}%
  \BibitemOpen
  \bibfield  {author} {\bibinfo {author} {\bibfnamefont {I.}~\bibnamefont
  {Regev}}, \bibinfo {author} {\bibfnamefont {J.}~\bibnamefont {Weber}},
  \bibinfo {author} {\bibfnamefont {C.}~\bibnamefont {Reichhardt}}, \bibinfo
  {author} {\bibfnamefont {K.~A.}\ \bibnamefont {Dahmen}}, \ and\ \bibinfo
  {author} {\bibfnamefont {T.}~\bibnamefont {Lookman}},\ }\href@noop {}
  {\bibfield  {journal} {\bibinfo  {journal} {Nature Communications}\ }\textbf
  {\bibinfo {volume} {6}},\ \bibinfo {pages} {8805} (\bibinfo {year}
  {2015})}\BibitemShut {NoStop}%
\bibitem [{\citenamefont {Das}\ \emph {et~al.}(2020)\citenamefont {Das},
  \citenamefont {Vinutha},\ and\ \citenamefont {Sastry}}]{sastry}%
  \BibitemOpen
  \bibfield  {author} {\bibinfo {author} {\bibfnamefont {P.}~\bibnamefont
  {Das}}, \bibinfo {author} {\bibfnamefont {H.~A.}\ \bibnamefont {Vinutha}}, \
  and\ \bibinfo {author} {\bibfnamefont {S.}~\bibnamefont {Sastry}},\
  }\href@noop {} {\bibfield  {journal} {\bibinfo  {journal} {PNAS}\ }\textbf
  {\bibinfo {volume} {117}},\ \bibinfo {pages} {10203} (\bibinfo {year}
  {2020})}\BibitemShut {NoStop}%
\bibitem [{\citenamefont {Schreck}\ \emph {et~al.}(2013)\citenamefont
  {Schreck}, \citenamefont {Hoy}, \citenamefont {Shattuck},\ and\ \citenamefont
  {O'Hern}}]{loop}%
  \BibitemOpen
  \bibfield  {author} {\bibinfo {author} {\bibfnamefont {C.~F.}\ \bibnamefont
  {Schreck}}, \bibinfo {author} {\bibfnamefont {R.~S.}\ \bibnamefont {Hoy}},
  \bibinfo {author} {\bibfnamefont {M.~D.}\ \bibnamefont {Shattuck}}, \ and\
  \bibinfo {author} {\bibfnamefont {C.~S.}\ \bibnamefont {O'Hern}},\
  }\href@noop {} {\bibfield  {journal} {\bibinfo  {journal} {Phys. Rev. E}\
  }\textbf {\bibinfo {volume} {88}},\ \bibinfo {pages} {052205} (\bibinfo
  {year} {2013})}\BibitemShut {NoStop}%
\bibitem [{\citenamefont {Royer}\ and\ \citenamefont
  {Chaikin}(2015)}]{cyclic_sand}%
  \BibitemOpen
  \bibfield  {author} {\bibinfo {author} {\bibfnamefont {J.~R.}\ \bibnamefont
  {Royer}}\ and\ \bibinfo {author} {\bibfnamefont {P.~M.}\ \bibnamefont
  {Chaikin}},\ }\href@noop {} {\bibfield  {journal} {\bibinfo  {journal}
  {PNAS}\ }\textbf {\bibinfo {volume} {112}},\ \bibinfo {pages} {49} (\bibinfo
  {year} {2015})}\BibitemShut {NoStop}%
\bibitem [{\citenamefont {VanderWerf}\ \emph {et~al.}(2018)\citenamefont
  {VanderWerf}, \citenamefont {Jin}, \citenamefont {Shattuck},\ and\
  \citenamefont {O'Hern}}]{Hypostatic}%
  \BibitemOpen
  \bibfield  {author} {\bibinfo {author} {\bibfnamefont {K.}~\bibnamefont
  {VanderWerf}}, \bibinfo {author} {\bibfnamefont {W.}~\bibnamefont {Jin}},
  \bibinfo {author} {\bibfnamefont {M.}~\bibnamefont {Shattuck}}, \ and\
  \bibinfo {author} {\bibfnamefont {C.~S.}\ \bibnamefont {O'Hern}},\
  }\href@noop {} {\bibfield  {journal} {\bibinfo  {journal} {Phys. Rev. E}\
  }\textbf {\bibinfo {volume} {97}},\ \bibinfo {pages} {012909} (\bibinfo
  {year} {2018})}\BibitemShut {NoStop}%
\bibitem [{\citenamefont {Silbert}\ \emph {et~al.}(2002)\citenamefont
  {Silbert}, \citenamefont {Ertas}, \citenamefont {Grest}, \citenamefont
  {Halsey},\ and\ \citenamefont {Levine}}]{silbertfriction}%
  \BibitemOpen
  \bibfield  {author} {\bibinfo {author} {\bibfnamefont {L.~E.}\ \bibnamefont
  {Silbert}}, \bibinfo {author} {\bibfnamefont {D.}~\bibnamefont {Ertas}},
  \bibinfo {author} {\bibfnamefont {G.~S.}\ \bibnamefont {Grest}}, \bibinfo
  {author} {\bibfnamefont {T.~C.}\ \bibnamefont {Halsey}}, \ and\ \bibinfo
  {author} {\bibfnamefont {D.}~\bibnamefont {Levine}},\ }\href@noop {}
  {\bibfield  {journal} {\bibinfo  {journal} {Phys. Rev. E}\ }\textbf {\bibinfo
  {volume} {65}},\ \bibinfo {pages} {031304} (\bibinfo {year}
  {2002})}\BibitemShut {NoStop}%
\end{thebibliography}%

\end{document}